\documentclass[journal]{IEEEtran}
\IEEEoverridecommandlockouts

\usepackage{graphicx,amsmath,amssymb,array,xcolor,booktabs,siunitx,eurosym,enumitem,balance}
\usepackage{tikz}
\usetikzlibrary{arrows.meta}
\usepackage{cite}
\usepackage{amsmath,amssymb,amsfonts}
\usepackage{amssymb}
\usepackage{graphicx}
\usepackage{textcomp}
\usepackage{xcolor}
\usepackage{subcaption}
\usepackage{float}
\usepackage[hidelinks]{hyperref}
\usepackage[acronym,nonumberlist,nopostdot,nomain,nogroupskip]{glossaries}

\def\x{{\mathbf x}}

\newtheorem{remark}{Remark}

\newcommand{\eqdef}{\mathrel{\triangleq}}

\newcommand{\Description}[1]{}

\definecolor{frameA}{HTML}{F03A3A}   
\definecolor{decision}{HTML}{19B35A} 
\definecolor{fillgray}{HTML}{F7F7F7}

\newacronym{lrfhss}{LR-FHSS}{Long Range-Frequency Hopping Spread Spectrum}
\newacronym{3d}{3D}{Three Dimensional}
\newacronym{6g}{6G}{Sixth Generation communication technology}
\newacronym{css}{CSS}{Chirp Spread Spectrum}
\newacronym{cc}{CC}{cross correlation}
\newacronym{ic}{IC}{inner correlation}
\newacronym{ac}{AC}{auto correlation}
\newacronym{scss}{SCSS}{Symmetry Chirp Spread Spectrum}
\newacronym[plural=UAVs,firstplural=Unmanned Aerial Vehicles (UAVs)]{uav}{UAV}{Unmanned Aerial Vehicle}
\newacronym[plural=LPWANs,firstplural=Low Power Wide Area Networks (LPWANs)]{lpwan}{LPWAN}{Low Power Wide Area Network}
\newacronym{isl}{ISL}{inter satellite link}
\newacronym{lora}{LoRa}{Long Range}
\newacronym{lte-m}{LTE-M}{Long Term Evolution for Machines}
\newacronym{iot}{IoT}{Internet of Things}
\newacronym{geo}{GEO}{Geostationary Orbit}
\newacronym[plural=HAPs,firstplural=High Altitude Platforms (HAPs)]{hap}{HAP}{High Altitude Platform}
\newacronym[plural=WANs,firstplural=Wide Area Networks (WANs)]{wan}{WAN}{Wide Area Network}
\newacronym{lorawan}{LoRaWAN}{Long Range Wide Area Network}
\newacronym{phy}{PHY}{physical layer}
\newacronym{gw}{GW}{Gateway}
\newacronym{nbiot}{NB-IoT}{NarrowBand-Internet of Things}
\newacronym{leo}{LEO}{Low Earth Orbit}
\newacronym[plural=NTNs,firstplural=Non-Terrestrial Networks (NTNs)]{ntn}{NTN}{Non-Terrestrial Network}
\newacronym[plural=GSs,firstplural=Ground Stations (GSs)]{gs}{GS}{Ground Station}
\newacronym[plural=BSs,firstplural=Basic Stations (BSs)]{bs}{BS}{Basic Station}
\newacronym[plural=FCs,firstplural=Frame Counters (FCs)]{fc}{FC}{Frame Counter}
\newacronym[plural=NSs,firstplural=Network Servers (NSs)]{ns}{NS}{Network Server}
\newacronym{plr}{PLR}{Packet Loss Rate}
\newacronym{cfo}{CFO}{Carrier Frequency Offset}
\newacronym{3gpp}{3GPP}{3rd Generation Partnership Project}

\newacronym{pdr}{PDR}{Packet Delivery Rate}
\newacronym{rmse}{RMSE}{Root Mean Square Error}
\newacronym[plural=ACKs, firstplural=acknowledgemnts (ACKs)]{ack}{ACK}{acknowledgment}
\newacronym{ip}{IP}{Internet Protocol}
\newacronym{rf}{RF}{Radio Frequency}
\newacronym[plural=ADRs,firstplural=Adaptive Data Rates (ADRs)]{adr}{ADR}{Adaptive Data Rate}
\newacronym{mcs}{MCS}{Modulation coding scheme}
\newacronym{mmtc}{mMTC}{Massive Machine Type Communications}
\newacronym{mac}{MAC}{Medium Access Control}
\newacronym{tcp}{TCP}{Transmission Control Protocol}
\newacronym{tls}{TLS}{Transport Layer Security}
\newacronym[plural=VBSs,firstplural=Virtual Basic Stations (VBSs)]{vbs}{VBS}{Virtual Basic Station}
\newacronym[plural=BRs,firstplural=ChirpStack Gateway Bridges (BRs)]{br}{BR}{ChirpStack Gateway Bridge}
\newacronym{zc}{ZC}{Zadoff-Chu}
\newacronym{dft}{DFT}{Discrete Foureir Transform}
\newacronym{lcm}{LCM}{Least Common Multiple}

\newacronym{mqtt}{MQTT}{Message Queuing Telemetry Transport}
\newacronym{otaa}{OTAA}{Over The Air Activation}
\newacronym{abp}{ABP}{Activation By Personalization}
\newacronym{ral}{RAL}{Radio Abstraction Layer}
\newacronym{mic}{MIC}{Message Integrity Code}
\newacronym[plural=QoSs,firstplural=Quality of Services (QoSs)]{qos}{QoS}{Quality of Service}
\newacronym{m2m}{M2M}{Machine-to-Machine}
\newacronym{rtt}{RTT}{Round-Trip Time}
\newacronym{hdop}{HDOP}{Horizontal Dilution of Precision}
\newacronym[plural=SFs,firstplural=Spreading Factors (SFs)]{sf}{SF}{Spreading Factor}
\newacronym[plural=hNSs,firstplural=Home Network Servers (HNSs)]{hns}{HNS}{Home Network Server}
\newacronym[plural=fNSs,firstplural=Forwarding Network Servers (FNSs)]{fns}{FNS}{Forwarding Network Server}
\newacronym[plural=SNSs,firstplural=Serving Network Servers (SNSs)]{sns}{SNS}{Serving Network Server}
\newacronym{pca}{PCA}{Point of Closest Approach}
\newacronym{qam}{QAM}{Quadrature Amplitude Modulation}
\newacronym{tof}{ToF}{Time of Flight}
\newacronym{aoa}{AoA}{Angle of Arrival}
\newacronym{iout}{IoUT}{Internet of Underwater Things}
\newacronym{em}{EM}{electromagnetic}
\newacronym{fscm}{FSCM}{Frequency Shift Chirp Modulation}
\newacronym{rssi}{RSSI}{Received Signal Strength Indicator}
\newacronym{per}{PER}{Packet Error Rate}
\newacronym{snr}{SNR}{Signal-to-Noise Ratio}
\newacronym{elf}{ELF}{Extremely Low Frequency}
\newacronym{vlf}{VLF}{Very Low Frequency}
\newacronym{hf}{HF}{High Frequency}
\newacronym{uhf}{UHF}{Ultra High Frequency}
\newacronym{uw2aw}{UW2AW}{underwater-to-above-water}
\newacronym{uw2uw}{UW2UW}{underwater-to-underwater}
\newacronym{cr}{CR}{Coding Rate}
\newacronym{fec}{FEC}{Forward Error Correction}
\newacronym{isi}{ISI}{inter-symbol inteference}
\newacronym{awgn}{AWGN}{Additive White Gaussian Noise}
\newacronym{sdf}{SDF}{Signal Doppler Frequency}
\newacronym{sdr}{SDR}{Software Defined Radio}
\newacronym{tdoa}{TDoA}{Time Difference of Arrival}
\newacronym{toa}{ToA}{Time of Arrival}
\newacronym{fei}{FEI}{frequency-error-indicator}
\newacronym{fdoa}{FDoA}{Frequency Difference of Arrival}
\newacronym{dpd}{DPD}{Direct Position Determination}
\newacronym{gnss}{GNSS}{Global Navigation Satellite System}
\newacronym{tcxo}{TCXO}{Temperature Controlled Crystal Oscillator}
\newacronym{ocxo}{OCXO}{Oven Controlled Crystal Oscillator}
\newacronym{ppm}{ppm}{parts-per-million}
\newacronym{pll}{PLL}{Phase-Locked Loop}
\newacronym{fft}{FFT}{Fast Fourier Transform}
\usetikzlibrary{arrows.meta,backgrounds,decorations.pathreplacing,calc}

\usepackage{algorithm,algpseudocode}
\usetikzlibrary{arrows.meta,positioning,shapes.geometric,fit,calc}

\def\BibTeX{{\rm B\kern-.05em{\sc i\kern-.025em b}\kern-.08em
    T\kern-.1667em\lower.7ex\hbox{E}\kern-.125emX}}
\begin{document}

\title{Sensing Fast and Distant Targets in Fractional Overspread Channels with Zak-OTFS and \\ Modular Arithmetic}

\author{
\IEEEauthorblockN{Andrea Novero, Mudasir Ahmad Sheikh, Danish Nisar, Saif Khan Mohammed,\\
 Michele Zorzi and Robert Calderbank}

\thanks{A. Novero, M. A. Sheikh, D. Nisar, and R. Calderbank are with the
Department of Electrical and Computer Engineering, Duke University, Durham,
NC, USA (e-mail: andrea.novero@duke.edu; mudasir.sheikh@duke.edu;
danish.nisar@duke.edu; robert.calderbank@duke.edu). A. Novero and M. Zorzi
are also with the Department of Information Engineering, University of
Padova, Italy (e-mail: andrea.novero@phd.unipd.it; michele.zorzi@unipd.it).
M. Zorzi is also with Wireless and More srl, Italy. S. K. Mohammed is with
Cohere Technologies Inc., San Jose, CA, USA, on extra-ordinary leave from the
Department of Electrical Engineering, Indian Institute of Technology Delhi,
New Delhi, India (e-mail: saifkmohammed@gmail.com). R. Calderbank is also
with Cohere Technologies Inc. A. Novero is the corresponding author (e-mail:
andrea@andreanovero.com).}
}

\maketitle
\begin{abstract}
Delay--Doppler (DD) modulations, such as Zak-transform-based orthogonal time frequency space (Zak-OTFS) modulation, represent the channel as a sparse set of delay and Doppler shifts, which makes them well suited for radar sensing. Sensing with a single DD frame, however, is unambiguous only when the channel spread factor, i.e., the product of the delay and Doppler spreads, is less then one. Long-range mobility scenarios, including drones, airplanes, low Earth orbit (LEO) satellites, and underwater acoustic links, typically violate this crystallization condition, so that a single frame aliases the target delays, Doppler shifts, or both. Existing methods add a time--frequency pilot to a conventional DD frame,
resolve the aliasing in only one dimension, and assume integer delays and
Doppler shifts. We propose a frame design that combines multiple DD bases
with different delay and Doppler periods, which together cover an
overspread channel with fractional delays and Doppler shifts, using the
same bandwidth and frame size. We show that modular arithmetic on the peak
positions uniquely recovers each target by pairing the peaks through their
common projection and their gains, and we derive the condition under which
this pairing is ambiguous. We implement the design with two superimposed Zak-OTFS frames and propose two receivers: dual-frame sensing (DS), which performs the joint detection directly, and DS with cross-pilot cancellation (DS-XC), which additionally cancels the cross-pilot interference, re-estimates the gains by joint least squares, and removes ghost targets. Both receivers maintain a detection rate close to one as the spread factor grows, and DS-XC also achieves the lowest false-alarm rate and channel normalized mean squared error among all the compared systems. In a 100~km drone scenario, DS-XC detects 98\% of the targets, compared with 81\% for the best baseline, with a complexity that grows near-linearly with the time--bandwidth product for a fixed number of targets.
\end{abstract}

\section{Introduction}

In the delay--Doppler (DD) domain, a doubly selective channel acts as a small set of delay and Doppler shifts, and each target appears directly as a peak at its own delay and Doppler. DD representations are therefore a natural choice for radar sensing. In high-mobility scenarios, orthogonal frequency division multiplexing (OFDM) suffers from inter-carrier interference and fast channel variations, whereas DD modulations resolve the paths and exploit the full time--frequency (TF) diversity. Several families of DD modulations have thus emerged, such as orthogonal time frequency space (OTFS) modulation~\cite{hadani2017otfs,raviteja2018interference,sheikh2024bayesian,sheikh2026variational}, orthogonal delay-Doppler division multiplexing (ODDM)~\cite{lin2022oddm}, and affine frequency division multiplexing (AFDM)~\cite{bemani2023afdm}, which are closely related through their pulse-train structure~\cite{mehrotra2026unification}. Among them, Zak-transform-based OTFS (Zak-OTFS)~\cite{mohammed2022otfs} provides a predictable input--output relation~\cite{mohammed2023predictability}.

The suitability of OTFS for radar was first demonstrated in~\cite{raviteja2019radar}, where the target range and velocity are estimated from the DD-domain response of an OTFS frame. Gaudio \textit{et al.}~\cite{gaudio2020otfs} derived maximum-likelihood estimators of the target parameters for OTFS and showed that their accuracy approaches the Cram\'er--Rao bound. The practical use of a single OTFS frame for sensing was recently demonstrated in the distributed sensing system for tracking (DISACT)~\cite{lin2026otfs}. For Zak-OTFS, the predictability of the input--output relation allows the DD response of the targets to be read directly from the received frame~\cite{mohammed2023predictability,dan2026zakotfs,MUbadahSpreading24}, and the radar cross-ambiguity function reduces to a set of fast Fourier transforms, with complexity near-linear in the time--bandwidth product~\cite{NishantDiscreteRadar25}. All these designs rely on the \emph{crystallization condition}, introduced for Zak-OTFS in~\cite{mohammed2023predictability}: the delay spread $\tau_{\max}$ and the Doppler spread $2\nu_{\max}$ of the channel must lie within the delay period $\tau_p$ and the Doppler period $\nu_p$ of the frame. Since $\tau_p\nu_p=1$, this is possible only if the spread factor $\varsigma \triangleq 2\tau_{\max}\nu_{\max}$ is below one. Otherwise, the channel is overspread, and the paths outside the periods are aliased.
\begin{figure*}[!h]
  \centering
  \includegraphics[width=\textwidth]{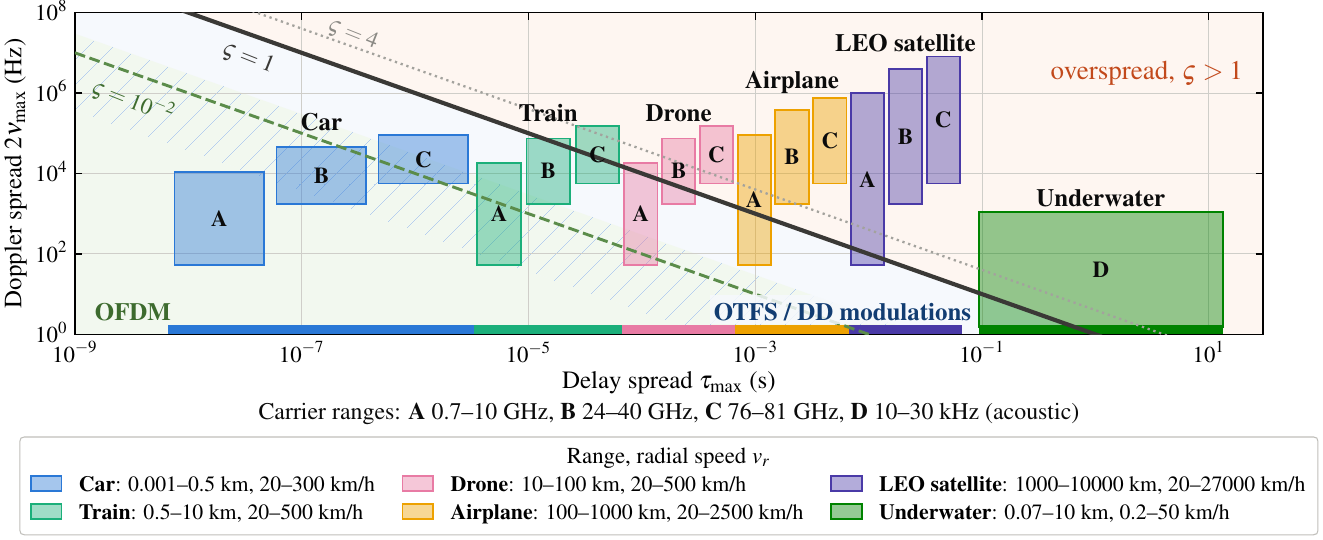}
  \caption{\textbf{Monostatic sensing capabilities of modulations.} Delay
  spread $\tau_{\max} = 2R_{\max}/c$ and Doppler spread
  $2\nu_{\max} = 4v_{r,\max}f_c/c$ of sensing scenarios in three RF ranges
  (A--C) and underwater acoustics (D, $c = 1500$~m/s), for targets with radial
  speed $v_r$. Each bar spans the Doppler range of a scenario in one carrier
  range; the bars of a scenario are drawn side by side, dividing the delay range of each scenario into three. OFDM sensing~\cite{sturm2011waveform} requires the
  cyclic prefix to cover the delay spread and the Doppler spread to be small
  with respect to the subcarrier spacing, i.e., $\varsigma \lesssim 10^{-2}$
  (hatched: transition); OTFS and the related DD modulations operate up to the
  crystallization limit $\varsigma = 1$. The dotted line is $\varsigma = 4$.}
  \label{fig:spread_map}
\end{figure*}

For monostatic sensing with maximum range $R_{\max}$ and targets with radial speed in $\pm v_{r,\max}$, the delay spread is $\tau_{\max}=2R_{\max}/c$ and the maximum Doppler shift is $\nu_{\max}=2v_{r,\max}f_c/c$, so the spread factor $\varsigma = 8R_{\max}v_{r,\max}f_c/c^2$ grows with range, speed, and carrier frequency. Fig.~\ref{fig:spread_map} shows the resulting sensing capabilities of OFDM and DD modulations on the DD plane, for typical mobility scenarios in three radio-frequency ranges and in underwater acoustics. OFDM sensing is limited to $\varsigma \lesssim 10^{-2}$~\cite{sturm2011waveform}, while DD modulations extend the operating region up to the crystallization limit $\varsigma=1$. Cars at short range remain underspread in every band, while trains become overspread at ranges of a few kilometers and millimeter-wave frequencies. Drones and airplanes at long range are overspread already below 10~GHz, low Earth orbit (LEO) satellites are overspread except at low radial speeds and low carriers, and moving underwater acoustic links are overspread except at short range and low speed. Overspread channels are therefore not a corner case, but the common situation in long-range mobility scenarios, where neither OFDM nor single-frame DD modulations can unambiguously recover the delay and Doppler of the targets.

In the OTFS literature, existing solutions for overspread channels keep a conventional DD frame and add a second, TF-domain pilot to resolve the aliasing in one dimension. For channels with large delay spread, Priya \textit{et al.}~\cite{priya2024delay} use a high-power DD pilot to obtain the aliased delays and the Doppler shifts. Rows whose pilot energy spreads along the Doppler axis are declared aliased, and the true delays are found by time-domain correlation with Doppler-matched dual chirps, followed by refinement steps. The Doppler shifts, however, are read directly from the DD grid, so the Doppler spread must remain within the Doppler period. For overspread Doppler, the same authors~\cite{priya2026doppler} combine the delays and aliased Doppler shifts given by a high-power DD pilot with the actual Doppler shifts given by the spectrum of the echoes of a cosine pilot, and pair them through their aliases and gain magnitudes. In this case, the delay spread must remain within the delay period. Both methods therefore cover only one overspread dimension, and both require pilot power ratios of $7$~dB~\cite{priya2024delay} and $28$~dB~\cite{priya2026doppler} between the DD pilot and the TF pilot. Moreover, they were designed for integer delays and Doppler shifts, and their estimation procedures can still produce ambiguous detections: paths sharing the same aliased delay require additional refinement steps~\cite{priya2024delay}, and paths with identical delay and aliased Doppler, or with close aliased fractional Doppler, are not reliably resolved~\cite{priya2026doppler}.

Overcoming the crystallization limit therefore requires a new frame design. We propose to use multiple bases that treat delay and Doppler differently and together cover an overspread channel. In this paper, we implement a simple form of the design, with two frames and DD point pilots, whose bases have swapped delay and Doppler dimensions. The proposed design is a general framework. As shown in~\cite{mehrotra2026unification,NishantDiscreteRadar25}, the bases of several modulations, such as Zak-OTFS~\cite{mohammed2022otfs}, ODDM~\cite{lin2022oddm}, orthogonal time sequency multiplexing (OTSM)~\cite{thaj2021otsm}, and AFDM~\cite{bemani2023afdm}, are obtained from the Zak-OTFS bases through unitary transformations, such as symplectic transformations or a generalization of the discrete affine Fourier transform, and any such transformation yields an equivalent system. A system demonstrated with Zak-OTFS therefore carries over to all the schemes obtained through these transformations. We choose Zak-OTFS as a simple implementation to demonstrate the new frame design.

The contributions of this paper are as follows: 

\begin{itemize}
    \item We propose a frame design for sensing overspread channels, built on multiple bases that treat delay and Doppler differently and jointly cover the channel spread. The design applies to any DD modulation and to the bases obtained from Zak-OTFS through unitary transformations.
    \item We develop a simple form of the design with two frames having DD point pilots and swapped delay and Doppler dimensions. We formulate its joint detection, which pairs the detections of the two frames through their common projection and their gains, and derive the condition under which the pairing is ambiguous. We further generalize the design to frames with non-divisible dimensions, which enlarge the unaliased region and reduce the pilot peak-to-average power ratio (PAPR).
    \item We implement the design with two Zak-OTFS frames superimposed in the same TF resource, and propose two receivers that handle fractional
    delays and Doppler shifts: dual-frame sensing (DS), and DS with cross-pilot cancellation (DS-XC), which additionally cancels the interference between the frames and removes ghost targets.
    \item We show that the proposed receivers outperform the single-frame design of~\cite{lin2026otfs} and the methods of~\cite{priya2024delay,priya2026doppler}, all implemented with Zak-OTFS bases, in detection rate, and that DS-XC also achieves the lowest false-alarm (FA) rate and channel normalized mean squared error (NMSE). In a long-range drone scenario, DS-XC achieves the highest detection rate and the lowest range and speed errors, with a complexity that grows near-linearly with the time--bandwidth product.
\end{itemize}

\section{Dual-Frame Sensing of Overspread Channels}
\label{sec:idea}

A DD frame observes the channel only modulo its delay and Doppler periods,
so a path outside these periods appears at a wrapped, i.e., aliased,
position. The key idea of this paper is that, by selecting bases with
different delay and Doppler periods, an overspread channel can be covered
unambiguously. In a simple implementation of this scheme, two DD frames with
different periods wrap the same path to different positions, and the pair of
wrapped positions identifies the path uniquely. This is the principle by which the residues of an integer modulo two
different numbers determine the integer over a range larger than either
number, and our design is an evolution of the approach by which multiple pulse
repetition frequency (PRF) radars resolve range
ambiguities~\cite{trunk1993range}. In this section, we develop the principle in its simplest form, with ideal filters, a point pilot, and integer delay and Doppler shifts, so that each path produces a single peak in one DD bin. Fractional shifts and practical filters do not change the principle, and are addressed in Sections~\ref{sec:system model} and~\ref{sec:proposed_detection}.

\subsection{The Single-Frame Limit}

Consider a Zak-OTFS frame of bandwidth $B$ and duration $T$, which carries $G = BT$ samples on a DD grid of $M$ delay bins and $N$ Doppler bins, with $G = MN$. The delay and Doppler resolutions are fixed by the bandwidth and the duration, $\Delta\tau = 1/B$ and $\Delta\nu = 1/T = B/G$, while the delay and Doppler periods depend on how $G$ is factorized:
\begin{equation}
  \tau_p = M\Delta\tau = \frac{M}{B}, \;
  \nu_p = N\Delta\nu = \frac{B}{M}, \;
  \tau_p\nu_p = 1 .
  \label{eq:periods}
\end{equation}
The factorization therefore does not change the resolution or the frame size, but only the shape of the unit-area region that the frame observes without aliasing: a frame with large $M$ is long in delay and narrow in Doppler, and a frame with small $M$ is the opposite.

A channel with delays in $[0,\tau_{\max})$ and Doppler shifts in $[-\nu_{\max},\nu_{\max}]$ fits within one frame only if $\tau_{\max} < \tau_p$ and $2\nu_{\max} < \nu_p$. Since $\tau_p\nu_p = 1$, this requires the spread factor $\varsigma = 2\tau_{\max}\nu_{\max}$ to be below one, which is the crystallization condition~\cite{mohammed2023predictability}. The same limit applies to all the related DD waveforms, such as MC-OTFS~\cite{hadani2017otfs,raviteja2018interference}, ODDM~\cite{lin2022oddm}, OTSM~\cite{thaj2021otsm}, and AFDM~\cite{bemani2023afdm}, which are instances of a single family of pulse-train waveforms~\cite{mehrotra2026unification}. For an overspread channel, $\varsigma > 1$, no factorization works: lengthening $\tau_p$ to cover the delay spread shortens $\nu_p$ by the same factor, and vice versa. For example, the DISACT system~\cite{lin2026otfs} uses $B = 100$~MHz at a carrier frequency of $4.5$~GHz, with $\tau_p \approx 8.3~\mu$s and $\nu_p \approx 120$~kHz. Its frame thus observes without aliasing a path length of only about $2.5$~km, i.e., a monostatic range of about $1.25$~km, for radial speeds up to about $\pm 2$~km/s. Targets beyond these limits are aliased, which is the case addressed in this paper.

\subsection{Two Frames With Swapped Dimensions}

The factorization of $G$ is a free design choice. The proposed system fixes $B$ and $G$ from the required delay and Doppler resolutions, and transmits two Zak-OTFS frames whose delay and Doppler dimensions are swapped:
\begin{equation}
  a \eqdef M_1 = N_2, \; b \eqdef M_2 = N_1, \; \rho \eqdef \frac{b}{a} > 1 .
  \label{eq:swap}
\end{equation}
Frame~1, with $\tau_{p_1} = a/B$ and $\nu_{p_1} = B/a$, is short in delay and wide in Doppler. Frame~2, with $\tau_{p_2} = b/B$ and $\nu_{p_2} = B/b$, is long in delay and narrow in Doppler. Both frames have the same resolutions and the same duration, and each still observes a region of unit area. Their combination, however, observes the delay range of frame~2 and the Doppler range of frame~1, i.e., a region of $b \times b$ bins with spread $\tau_{p_2}\nu_{p_1} = \rho$. The two fundamental regions overlap on a square of $a \times a$ bins, with spread $\tau_{p_1}\nu_{p_2} = 1/\rho$, which is drawn as a blue dotted square in Fig.~\ref{fig:overspread}. In the example of Fig.~\ref{fig:overspread}, $a = 5$ and $b = 15$, so that $G = 75$ and the two frames jointly cover a spread $\rho = 3$.

\begin{figure*}[htbp]
  \centering
  \begin{tikzpicture}[
  scale=0.8,
  dimarrow/.style={{Latex[length=2mm]}-{Latex[length=2mm]}, line width=0.8pt},
  resarrow/.style={-{Latex[length=1.6mm]}, line width=0.6pt},
  extline/.style={line width=0.3pt, black!55},
  tick/.style={line width=0.7pt},
  gridc/.style={black!18, line width=0.3pt},
  det/.style={draw=#1, fill=white, line width=1.1pt},
  dettwo/.pic={\draw[fill=white, line width=1.1pt, pic actions] (-0.17,-0.17) rectangle (0.17,0.17);},
  bigcross/.pic={\draw[line width=1.4pt, line cap=round, pic actions]
                   (-0.13,-0.13) -- (0.13,0.13) (-0.13,0.13) -- (0.13,-0.13);},
  proj/.pic={\fill[pic actions] (90:0.12) -- (210:0.12) -- (330:0.12) -- cycle;},
  projdown/.pic={\fill[pic actions] (270:0.12) -- (30:0.12) -- (150:0.12) -- cycle;},
  ghost/.pic={\draw[line width=1.4pt, line cap=round, black!50, scale=1.4]
                   (-0.13,-0.13) -- (0.13,0.13) (-0.13,0.13) -- (0.13,-0.13);
              \draw[black!45, line width=0.6pt, dash pattern=on 1.5pt off 1.2pt] (0,0) circle (0.26);},
  panel/.style={font=\small},
]
\def\s{1}
\def\L{3*\s}
\pgfmathsetmacro{\tS}{2*\s/5}   
\pgfmathsetmacro{\Tw}{2*\s}     
\pgfmathsetmacro{\Wd}{2*\L}     

\newcommand{\ddlayout}[1]{%
  \fill[white] (0,-\L) rectangle (\Wd,\L);
  \draw[gridc, step=\tS, shift={(0,-\L)}] (0,0) grid ({\Wd+0.001},{\Wd+0.001});
  \draw[frameA, line width=1.3pt] (\Tw,\s) -- (\Tw,\L) -- (0,\L) -- (0,-\L) -- (\Tw,-\L) -- (\Tw,-\s);
  \draw[black, line width=1.3pt] (\Tw,\s) -- (\Wd,\s) -- (\Wd,-\s) -- (\Tw,-\s);
  \pgfmathsetmacro{\tA}{0}\pgfmathsetmacro{\tB}{\tS}\pgfmathsetmacro{\tE}{\Tw+0.01}
  \foreach \x in {\tA,\tB,...,\tE} \draw[tick,frameA] (\x,\L) -- (\x,\L-0.13);
  \pgfmathsetmacro{\tA}{\s}\pgfmathsetmacro{\tB}{\s+\tS}\pgfmathsetmacro{\tE}{\L+0.01}
  \foreach \y in {\tA,\tB,...,\tE} \draw[tick,frameA] (0,\y) -- (0.13,\y);
  \pgfmathsetmacro{\tA}{-\L}\pgfmathsetmacro{\tB}{-\L+\tS}\pgfmathsetmacro{\tE}{-\s+0.01}
  \foreach \y in {\tA,\tB,...,\tE} \draw[tick,frameA] (0,\y) -- (0.13,\y);
  \pgfmathsetmacro{\tA}{\Tw}\pgfmathsetmacro{\tB}{\Tw+\tS}\pgfmathsetmacro{\tE}{\Wd+0.01}
  \foreach \x in {\tA,\tB,...,\tE} \draw[tick] (\x,\s) -- (\x,\s-0.13);
  \pgfmathsetmacro{\tA}{-\s}\pgfmathsetmacro{\tB}{-\s+\tS}\pgfmathsetmacro{\tE}{\s+0.01}
  \foreach \y in {\tA,\tB,...,\tE} \draw[tick] (\Wd,\y) -- (\Wd-0.13,\y);
  \draw[#1, line width=2.2pt] (0,-\s) rectangle (\Tw,\s);
  \draw[blue!75!black, line width=1.2pt, dotted] (0,-\s) rectangle (\Tw,\s);
  \draw[frameA, line width=1.3pt] (0,-\s) -- (0,\s);
}

\definecolor{tgtA}{HTML}{E69F00}   
\definecolor{tgtB}{HTML}{8E44AD}   
\definecolor{tgtC}{HTML}{009E73}   
\definecolor{tgtH}{HTML}{0072B2}   
\definecolor{tgtP}{HTML}{CC79A7}   
\definecolor{tgtQ}{HTML}{A0522D}   

\begin{scope}
\ddlayout{white}

\pic[draw=tgtH, scale=1.25] at (0.6,0) {dettwo};       
\draw[det=tgtH] (0.6,0) circle (0.2);                  
\pic[tgtH!70, scale=1.4] at (0.6,0) {bigcross};        
\pic[tgtH] at (0.6,0) {proj};                          

\draw[extline] (\Tw+2*\tS,-\L) -- (\Tw+2*\tS,-\L-0.45);
\draw[extline] (\Tw+3*\tS,-\L) -- (\Tw+3*\tS,-\L-0.45);
\draw[resarrow] (\Tw+2*\tS-0.4,-\L-0.3) -- (\Tw+2*\tS,-\L-0.3);
\draw[resarrow] (\Tw+3*\tS+0.4,-\L-0.3) -- (\Tw+3*\tS,-\L-0.3);
\node[below] at (\Tw+2.5*\tS,-\L-0.35) {$\Delta\tau=\frac{1}{B}$};
\draw[extline] (\Wd,\L-2*\tS) -- (\Wd+0.45,\L-2*\tS);
\draw[extline] (\Wd,\L-3*\tS) -- (\Wd+0.45,\L-3*\tS);
\draw[resarrow] (\Wd+0.3,\L-2*\tS+0.3) -- (\Wd+0.3,\L-2*\tS);
\draw[resarrow] (\Wd+0.3,\L-3*\tS-0.3) -- (\Wd+0.3,\L-3*\tS);
\node[rotate=90] at (\Wd+0.7,\L-2.5*\tS) {$\Delta\nu=\frac{B}{G}$};

\draw[extline] (0,-\L) -- (0,-\L-1.55);
\draw[extline] (\Tw,-\L) -- (\Tw,-\L-0.6);
\draw[dimarrow, frameA] (0,-\L-0.45) -- (\Tw,-\L-0.45);
\node[frameA, below] at (\s,-\L-0.5) {$\tau_{p1}$};
\draw[extline] (\Wd,-\s) -- (\Wd,-\L-1.55);
\draw[dimarrow] (0,-\L-1.4) -- (\Wd,-\L-1.4);
\node[below] at (\L,-\L-1.45) {$\tau_{p2}$};
\draw[extline] (\Wd,\s)  -- (\Wd+0.6,\s);
\draw[extline] (\Wd,-\s) -- (\Wd+0.6,-\s);
\draw[dimarrow] (\Wd+0.45,-\s) -- (\Wd+0.45,\s);
\node[right] at (\Wd+0.5,0) {$\nu_{p2}$};
\draw[extline] (\Tw,\L)  -- (\Wd+1.55,\L);
\draw[extline] (\Tw,-\L) -- (\Wd+1.55,-\L);
\draw[dimarrow, frameA] (\Wd+1.4,-\L) -- (\Wd+1.4,\L);
\node[frameA, right] at (\Wd+1.45,0) {$\nu_{p1}$};

\node[panel] at (\L,-\L-3.7) {(a)};
\end{scope}

\begin{scope}[shift={(9.4,0)}]
\ddlayout{white}

\draw[det=tgtA] (1.0,-2.0) circle (0.2);
\pic[draw=tgtA, scale=1.25] at (1.0, 0.0) {dettwo};
\pic[tgtA!70, scale=1.4] at (1.0,-2.0) {bigcross};
\pic[tgtA] at (1.0, 0.0) {proj};
\pic[draw=tgtB, scale=1.25] at (3.4,-0.4) {dettwo};
\draw[det=tgtB] (1.4,-0.4) circle (0.2);
\pic[tgtB!70, scale=1.4] at (3.4,-0.4) {bigcross};
\pic[tgtB] at (1.4,-0.4) {proj};
\pic[draw=tgtC, scale=1.25] at (4.6, 0.4) {dettwo};   
\draw[det=tgtC] (0.6, 2.4) circle (0.2);              
\pic[tgtC!70, scale=1.4] at (4.6, 2.4) {bigcross};
\pic[tgtC] at (0.6, 0.4) {proj};

\node[panel] at (\L,-\L-3.7) {(b)};
\end{scope}

\begin{scope}[shift={(16.6,0)}]
\ddlayout{white}

\pic[draw=tgtP, scale=1.25] at (4.6, 0.4) {dettwo};
\draw[det=tgtP] (0.6, 2.4) circle (0.2);
\pic[draw=tgtQ, scale=1.25] at (2.6, 0.4) {dettwo};
\draw[det=tgtQ] (0.6,-1.6) circle (0.2);
\pic[tgtP!70, scale=1.4] at (4.6, 2.4) {bigcross};
\pic[tgtQ!70, scale=1.4] at (2.6,-1.6) {bigcross};
\pic at (2.6, 2.4) {ghost};
\pic at (4.6,-1.6) {ghost};
\pic[tgtP, scale=1.3] at (0.6,0.4) {proj};
\pic[tgtQ, scale=1.3] at (0.6,0.4) {projdown};

\node[panel] at (\L,-\L-3.7) {(c)};
\end{scope}

\begin{scope}[shift={(9.4,-\L-0.45)}, font=\footnotesize]
  \draw[black!60, line width=0.4pt, fill=white] (0,0) rectangle (13.2,-2.75);
  \draw[det=black!70] (0.4,-0.4) circle (0.15);
  \node[anchor=west] at (0.9,-0.4) {Detection, frame~1};
  \pic[draw=black!70] at (0.4,-1.05) {dettwo};
  \node[anchor=west] at (0.9,-1.05) {Detection, frame~2};
  \pic[black!70] at (0.3,-1.7) {proj};
  \node[inner sep=0pt, anchor=base, font=\large] at (0.52,-1.78) {,};
  \pic[black!70] at (0.68,-1.65) {projdown};
  \node[anchor=west] at (0.9,-1.7) {Projection onto overlap square};
  \pic[black!55] at (7.0,-0.4) {bigcross};
  \node[anchor=west] at (7.5,-0.4) {Real target (joint detection)};
  \pic[scale=0.75] at (7.0,-1.05) {ghost};
  \node[anchor=west] at (7.5,-1.05) {Ghost target};
  \fill[tgtH] (0.3,-2.45) rectangle (0.5,-2.25);  \node[anchor=west] at (0.55,-2.35) {Hyp.~1};
  \fill[tgtA] (2.0,-2.45) rectangle (2.2,-2.25);  \node[anchor=west] at (2.25,-2.35) {Hyp.~2};
  \fill[tgtB] (3.7,-2.45) rectangle (3.9,-2.25);  \node[anchor=west] at (3.95,-2.35) {Hyp.~3};
  \fill[tgtC] (5.4,-2.45) rectangle (5.6,-2.25);  \node[anchor=west] at (5.65,-2.35) {Hyp.~4};
  \fill[tgtP] (7.3,-2.45) rectangle (7.5,-2.25);
  \fill[tgtQ] (7.55,-2.45) rectangle (7.75,-2.25);\node[anchor=west] at (7.8,-2.35) {Ambiguous targets};
\end{scope}
\end{tikzpicture}
  \caption{Joint detection with two swapped Zak-OTFS frames; frame~1 is in red, frame~2 in black, and the overlap square is the blue dotted square. (a)~A target in the overlap square is detected at its true position by both frames. (b)~Targets inside frame~1 only, inside frame~2 only, and outside both. Each target is recovered by pairing its two detections through their common projection onto the overlap square. (c)~Ambiguous detection: two targets whose delay and Doppler differences are multiples of $\tau_{p_1}$ and $\nu_{p_2}$ have the same projection. If their amplitudes are indistinguishable, both pairings are consistent, and two ghost targets appear along with the two real targets.}
  \label{fig:overspread}
\end{figure*}

\subsection{Locating a Target From Two Aliased Peaks}

The DD-domain Zak-OTFS signal is quasi-periodic: it is periodic in Doppler with period $\nu_p$, and periodic up to a phase factor in delay with period $\tau_p$. The magnitude of the received response is therefore periodic in both dimensions, and each frame observes a path only through its wrapped position. Let $\langle \nu \rangle_{\nu_p}$ denote $\nu$ wrapped into $[-\nu_p/2, \nu_p/2)$. Frame~$f$ detects path~$i$, at $(\tau_i,\nu_i)$, as a peak at
\begin{equation}
  (\hat\tau^{(f)}_i, \hat\nu^{(f)}_i) = \big(\tau_i \bmod \tau_{pf},\
  \langle \nu_i \rangle_{\nu_{pf}}\big),
  \label{eq:wrapped_peak}
\end{equation}
which coincides with the true position only if the path lies in the
fundamental region $\mathcal{R}_f = [0, \tau_{pf}) \times
[-\nu_{pf}/2, \nu_{pf}/2)$ of the frame.

Consider a path within the region covered by the two frames, i.e., with
$\tau_i < \tau_{p2}$ and $|\nu_i| < \nu_{p1}/2$. Frame~2 then always
measures its delay correctly, and frame~1 always measures its Doppler
correctly: each frame supplies exactly the coordinate that the other may
alias. The coordinates of the path are therefore recovered with a single
rule,
\begin{equation}
  (\hat\tau_i, \hat\nu_i) = (\hat\tau^{(2)}_i, \hat\nu^{(1)}_i),
  \label{eq:reconstruction}
\end{equation}
wherever it lies. Depending on its position relative to the two fundamental regions, four cases arise, which are summarized in Table~\ref{tab:hypotheses} and shown in Figs.~\ref{fig:overspread}a and~\ref{fig:overspread}b. The receiver does not need to know which case holds, but the case can be read from the relation between the two peaks: they coincide when the target lies in the overlap square, they share the delay or the Doppler when it lies in only one fundamental region, and they differ in both coordinates when it lies outside both.

\begin{table}[!t]
  \centering
  \caption{Detection cases for a target within the region covered by the two frames.}
  \label{tab:hypotheses}
  \begin{tabular}{lccc}
    \hline
    Target location & Frame~1 peak & Frame~2 peak & Peaks share \\
    \hline
    $\mathcal{R}_1 \cap \mathcal{R}_2$ & true & true & both coordinates \\
    $\mathcal{R}_1$ only & true & Doppler wrapped & delay \\
    $\mathcal{R}_2$ only & delay wrapped & true & Doppler \\
    outside both & delay wrapped & Doppler wrapped & neither \\
    \hline
  \end{tabular}
\end{table}

With several targets, the receiver must also decide which peak of frame~1 belongs with which peak of frame~2 before applying~\eqref{eq:reconstruction}. The overlap square provides the common reference for this decision, since it is the only part of the DD plane that both frames observe in the same way. Projecting a peak onto it, i.e., reducing its delay modulo $\tau_{p1}$ and
wrapping its Doppler to $[-\nu_{p2}/2, \nu_{p2}/2)$, removes the aliasing
that differs between the two frames. By~\eqref{eq:wrapped_peak}, the two
peaks of path~$i$ both project onto
$(\tau_i \bmod \tau_{p1}, \langle \nu_i \rangle_{\nu_{p2}})$, i.e.,
\begin{equation}
  \hat\tau^{(2)}_i \bmod \tau_{p1} = \hat\tau^{(1)}_i, \qquad
  \langle \hat\nu^{(1)}_i \rangle_{\nu_{p2}} = \hat\nu^{(2)}_i .
  \label{eq:consistency}
\end{equation}
A pair of peaks that violates~\eqref{eq:consistency} cannot belong to the same target and is discarded. Moreover, the two peaks of the same target are scaled by the same path gain, so their amplitudes provide a second criterion, which separates targets that share a projection but have different gains.

\subsection{Ambiguous Pairings and Ghost Targets}

Pairing fails only when two paths $i$ and $i'$ project onto the same bin of
the overlap square, i.e., when
\begin{equation}
  \tau_i - \tau_{i'} \equiv 0 \pmod{\tau_{p1}}, \qquad
  \nu_i - \nu_{i'} \equiv 0 \pmod{\nu_{p2}},
  \label{eq:ambiguity}
\end{equation}
and their amplitudes are also indistinguishable. Both the true and the
swapped pairings then satisfy~\eqref{eq:consistency}, and the receiver
cannot distinguish them. The two true targets are then accompanied by two
ghost targets at $(\tau_i,\nu_{i'})$ and $(\tau_{i'},\nu_i)$, as shown in
Fig.~\ref{fig:overspread}c.

For $P$ targets placed uniformly at random over the covered region, the projection of each target is uniform over the $a^2$ bins of the overlap square, so two given targets collide with probability $1/a^2$. By the union bound over the $P(P-1)/2$ pairs, the probability that at least one pair is ambiguous satisfies
\begin{equation}
  p_{\mathrm{ghost}} \le \frac{P(P-1)}{2a^2},
  \label{eq:prob_basic}
\end{equation}
e.g., $p_{\mathrm{ghost}} \le 0.03$ for $P = 16$ and $a = 64$. Since $a = \sqrt{G/\rho}$, covering a larger spread with the same frame size shrinks the overlap square and makes ghosts more likely. This trade-off motivates the generalization of the next section.

\section{Generalization: Frames With Non-Divisible Dimensions}
\label{sec:general}

In the design of Fig.~\ref{fig:overspread}, $M_2$ is a multiple of $M_1$, so the delay period of frame~2 contains an integer number of delay periods of frame~1, and the two frames together cover at most the larger period in each dimension. Modular arithmetic suggests a better choice: two frames whose dimensions are not multiples of each other jointly cover a region larger than either frame. For example, knowing an integer modulo $8$ and modulo $12$ determines it
modulo $24$, a range larger than both moduli. As we show next, they also enlarge the region that both frames observe without aliasing. The two frames only need to share $B$ and $G$: their delay dimensions $M_1$ and $M_2$ can be any two divisors of $G$, with $N_f = G/M_f$.

\subsection{Joint Coverage}

Let $d = \gcd(M_1, M_2)$ and $L = \mathrm{lcm}(M_1, M_2) = M_1M_2/d$, and let $k = \tau/\Delta\tau$ and $l = \nu/\Delta\nu$ be the delay and Doppler indices of a path. By~\eqref{eq:wrapped_peak}, frame~$f$ observes the path at $k^{(f)} = k \bmod M_f$ and $l^{(f)} = \langle l \rangle_{N_f}$.
Indeed, the pair of residues $(k \bmod M_1, k \bmod M_2)$ repeats with period $L$, and within one period it takes $L = M_1M_2/d$ distinct values, which is exactly the number of consistent residue pairs, so each consistent pair corresponds to a single index $k \in\{0, \ldots, L-1\}$; the same argument holds for the Doppler index.

Since $\gcd(N_1, N_2) = G/L$, the two observations of the same path satisfy
\begin{equation}
  k^{(1)} \equiv k^{(2)} \pmod{d}, \qquad l^{(1)} \equiv l^{(2)} \pmod{G/L},
  \label{eq:crt_consistency}
\end{equation}
i.e., they project onto the same bin of a projection grid of $d \times G/L$ bins, which generalizes~\eqref{eq:consistency}. For a consistent pair, there is a unique solution of
\begin{equation}
  k = k^{(1)} + xM_1 = k^{(2)} + yM_2, \qquad l = l^{(1)} + x'N_1 = l^{(2)} + y'N_2,
  \label{eq:crt}
\end{equation}
with $x, y, x', y' \in \mathbb{Z}$, for $k \in [0, L)$ and $l \in [-G/(2d), G/(2d))$, since $\mathrm{lcm}(N_1, N_2) = G/d$. Eq.~\eqref{eq:crt} replaces the reconstruction rule~\eqref{eq:reconstruction}, and the two frames cover $L \times G/d$ bins, i.e., a spread
\begin{equation}
  \rho = \frac{L}{B}\cdot\frac{B}{d} = \frac{M_1 M_2}{d^2}.
  \label{eq:crt_spread}
\end{equation}
For the swapped frames of~\eqref{eq:swap}, $d = a$ and $L = b$, and~\eqref{eq:crt_consistency}--\eqref{eq:crt_spread} reduce to~\eqref{eq:consistency}, \eqref{eq:reconstruction}, and $\rho = b/a$, respectively. Since $L$ divides $G$, the covered delays $L/B$ never exceed the frame duration $T = G/B$.

\subsection{Unaliased Region}

Targets in the overlap $\mathcal{R}_1 \cap \mathcal{R}_2$ of the two fundamental regions, which has $\min(M_1, M_2) \times \min(N_1, N_2)$ bins, are detected at their true position by both frames without relying on~\eqref{eq:crt}. In the swapped design of Fig.~\ref{fig:overspread}, this overlap coincides with the projection grid, so it shrinks as $a = \sqrt{G/\rho}$ when the covered spread grows. With non-divisible dimensions, the overlap is decoupled from the projection grid. Since $\min_f M_f \ge d$ and $\min_f N_f \ge G/L$, it is always at least as large as the projection grid, and it is typically much larger, so the same spread is covered with a larger region free of aliasing.

\subsection{Ambiguity}

Two targets are ambiguous when they project onto the same bin of the projection grid and have indistinguishable amplitudes. For $P$ targets placed uniformly at random, the union bound over the $dG/L$ bins of the grid gives
\begin{equation}
  p_{\mathrm{ghost}} \le \frac{P(P-1)}{2}\cdot\frac{L}{dG}
  = \frac{P(P-1)}{2}\cdot\frac{\rho}{G},
  \label{eq:crt_ghost}
\end{equation}
which reduces to~\eqref{eq:prob_basic} for swapped frames, with $G = \rho a^2$. For a given frame size, the ambiguity therefore depends only on the covered spread, and not on the choice of $M_1$ and $M_2$. Covering a larger spread costs ambiguity regardless of how the frames are shaped, while the choice of dimensions determines the unaliased region and the PAPR.
\begin{figure}[!t]
  \centering
  \resizebox{\columnwidth}{!}{\begin{tikzpicture}[
  scale=1,
  every node/.style={font=\fontsize{16}{19}\selectfont},
  dimarrow/.style={{Latex[length=3mm]}-{Latex[length=3mm]}, line width=1pt},
  extline/.style={line width=0.4pt, black!55},
  gridc/.style={black!15, line width=0.25pt},
  det/.style={draw=#1, fill=white, line width=1.4pt},
  dettwo/.pic={\draw[fill=white, line width=1.4pt, pic actions] (-0.16,-0.16) rectangle (0.16,0.16);},
  bigcross/.pic={\draw[line width=2pt, line cap=round, pic actions]
                   (-0.2,-0.2) -- (0.2,0.2) (-0.2,0.2) -- (0.2,-0.2);},
  proj/.pic={\fill[pic actions] (90:0.17) -- (210:0.17) -- (330:0.17) -- cycle;},
  cand/.style={line width=0.7pt, dash pattern=on 2pt off 2pt},
]
\def\s{0.3}
\definecolor{tgtC}{HTML}{009E73}
\pgfmathsetmacro{\Lx}{24*\s}  
\pgfmathsetmacro{\Ly}{24*\s}  
\pgfmathsetmacro{\tx}{21.5*\s}\pgfmathsetmacro{\ty}{15.5*\s}
\def\ca{0.5}\def\cb{1.1}   

\fill[white] (0,0) rectangle (\Lx,\Ly);
\draw[gridc, step=\s] (0,0) grid ({\Lx+0.001},{\Ly+0.001});
\draw[black!70, line width=1.6pt, dash pattern=on 6pt off 3pt] (0,0) rectangle (\Lx,\Ly);
\draw[frameA, line width=1.6pt] (0,0) rectangle (8*\s,12*\s);
\draw[black, line width=1.6pt] (0,0) rectangle (12*\s,8*\s);
\draw[blue!75!black, line width=1.5pt, dotted] (0,0) rectangle (4*\s,4*\s);
\node[text=blue!75!black, anchor=south, inner sep=1pt] at (2*\s,4*\s) {$d$};
\node[text=blue!75!black, anchor=west, inner sep=1pt] at (4*\s,1.5*\s) {$G/L$};

\foreach \k in {5,13,21} \draw[det=frameA] ({(\k+0.5)*\s},-\ca) circle (0.14);
\foreach \k in {9,21}    \pic[draw=black] at ({(\k+0.5)*\s},-\cb) {dettwo};
\node[anchor=east, frameA] at (-0.1,-\ca) {$k_1 + xM_1$};
\node[anchor=east] at (-0.1,-\cb) {$k_2 + yM_2$};
\foreach \l in {3,15}    \draw[det=frameA] (\Lx+\ca,{(\l+0.5)*\s}) circle (0.14);
\foreach \l in {7,15,23} \pic[draw=black] at (\Lx+\cb,{(\l+0.5)*\s}) {dettwo};
\node[frameA, rotate=90, anchor=east] at (\Lx+\ca,-0.15) {$l_1 + xN_1$};
\node[rotate=90, anchor=east] at (\Lx+\cb,-0.15) {$l_2 + yN_2$};
\draw[cand, tgtC] (\tx,-\cb-0.2) -- (\tx,\ty);
\draw[cand, tgtC] (\Lx+\cb+0.2,\ty) -- (\tx,\ty);

\draw[det=tgtC] (5.5*\s,3.5*\s) circle (0.19);
\pic[draw=tgtC] at (9.5*\s,7.5*\s) {dettwo};
\pic[tgtC] at (1.5*\s,3.5*\s) {proj};
\pic[tgtC!70] at (\tx,\ty) {bigcross};

\draw[extline] (0,0) -- (0,-2.9);
\draw[extline] (8*\s,0) -- (8*\s,-1.9);
\draw[extline] (12*\s,0) -- (12*\s,-2.4);
\draw[extline] (\Lx,0) -- (\Lx,-2.9);
\draw[dimarrow, frameA] (0,-1.75) -- (8*\s,-1.75);
\draw[dimarrow] (0,-2.25) -- (12*\s,-2.25);
\draw[dimarrow, black!70] (0,-2.75) -- (\Lx,-2.75);
\node[frameA, anchor=west] at (8*\s+0.05,-1.75) {$\tau_{p1}$};
\node[anchor=west] at (12*\s+0.05,-2.25) {$\tau_{p2}$};
\node[anchor=north] at (\Lx/2,-2.8) {$L/B$};
\node at (\Lx/2,-3.8) {Delay};
\draw[extline] (0,8*\s) -- (-0.65,8*\s);
\draw[extline] (0,12*\s) -- (-1.2,12*\s);
\draw[extline] (0,\Ly) -- (-1.75,\Ly);
\draw[extline] (0,0) -- (-1.75,0);
\draw[dimarrow] (-0.5,0) -- (-0.5,8*\s);
\draw[dimarrow, frameA] (-1.05,0) -- (-1.05,12*\s);
\draw[dimarrow, black!70] (-1.6,0) -- (-1.6,\Ly);
\node[anchor=south, inner sep=1pt] at (-0.5,8*\s) {$\nu_{p2}$};
\node[frameA, anchor=south, inner sep=1pt] at (-1.05,12*\s) {$\nu_{p1}$};
\node[anchor=south, inner sep=1pt] at (-1.6,\Ly) {$B/d$};
\node[rotate=90] at (-2.25,\Ly/2) {Doppler};

\begin{scope}[shift={(0.1,\Ly-0.1)}, every node/.style={font=\fontsize{10}{12}\selectfont, anchor=west, inner sep=0.5pt}]
  \def\dy{0.41}
  \fill[white, opacity=0.92] (0,0) rectangle (7.0,{-5*\dy-0.08});
  \draw[black!40, line width=0.4pt] (0,0) rectangle (7.0,{-5*\dy-0.08});
  \draw[frameA, line width=1.3pt] (0.1,-0.24) -- (0.5,-0.24); \node at (0.58,-0.24) {Frame 1};
  \draw[black, line width=1.3pt] (0.1,{-0.24-\dy}) -- (0.5,{-0.24-\dy}); \node at (0.58,{-0.24-\dy}) {Frame 2};
  \draw[black!70, line width=1.3pt, dash pattern=on 4pt off 2pt] (0.1,{-0.24-2*\dy}) -- (0.5,{-0.24-2*\dy}); \node at (0.58,{-0.24-2.05*\dy}) {Covered region};
  \draw[blue!75!black, line width=1.2pt, dotted] (0.1,{-0.24-3*\dy}) -- (0.5,{-0.24-3*\dy}); \node at (0.58,{-0.24-3*\dy}) {Projection cell};
  \pic[tgtC!70, scale=0.7] at (0.3,{-0.24-4*\dy}) {bigcross}; \node at (0.58,{-0.24-4*\dy}) {True target};
  \draw[det=tgtC, line width=1.1pt] (3.6,-0.24) circle (0.13); \node at (3.82,-0.24) {Frame 1 detection};
  \pic[draw=tgtC, scale=0.75] at (3.6,{-0.24-\dy}) {dettwo}; \node at (3.82,{-0.24-\dy}) {Frame 2 detection};
  \pic[tgtC, scale=0.8] at (3.6,{-0.24-2.05*\dy}) {proj}; \node at (3.82,{-0.24-2.05*\dy}) {Common projection};
  \draw[det=frameA, line width=1.1pt] (3.6,{-0.24-3*\dy}) circle (0.1); \node at (3.82,{-0.24-3*\dy}) {Frame 1 candidates};
  \pic[draw=black, scale=0.75] at (3.6,{-0.24-4*\dy}) {dettwo}; \node at (3.82,{-0.24-4*\dy}) {Frame 2 candidates};
\end{scope}
\end{tikzpicture}}
\caption{Two frames with non-divisible dimensions, $G = 96$: frame~1 with $(M_1, N_1) = (8, 12)$ and frame~2 with $(M_2, N_2) = (12, 8)$, which jointly cover $24 \times 24$ bins. A target outside both frames is observed at wrapped positions with a common projection in the $4 \times 4$ projection grid, and the candidate positions of the two frames coincide only at its true position. For
  readability, the Doppler index starts from zero.}
  \label{fig:crt}
\end{figure}
\subsection{Example}
Fig.~\ref{fig:crt} illustrates the reconstruction for $G = 96$, with frame~1 of size $(M_1, N_1) = (8, 12)$ and frame~2 of size $(M_2, N_2) = (12, 8)$, so that $d = 4$ and $L = 24$. Consider a target at $(k, l) = (21, 15)$, which lies outside both fundamental regions. Frame~1 observes it at $(21 \bmod 8,\, 15 \bmod 12) = (5, 3)$, and frame~2 at $(21 \bmod 12,\, 15 \bmod 8) = (9, 7)$. Neither observation is correct, but both project onto the same bin $(1, 3)$ of the $4 \times 4$ projection grid, which identifies them as the same target. Each observation then defines a set of candidate positions, spaced by the period of its frame. Along the delay axis, frame~1 gives the candidates $\{5, 13, 21\}$ and frame~2 gives $\{9, 21\}$, and along the Doppler axis, frame~1 gives $\{3, 15\}$ and frame~2 gives $\{7, 15, 23\}$. The only position that both frames agree on is the true one, $(21, 15)$.
The divisible swapped design covers the same $24 \times 24$ bins, i.e., $\rho = 6$, only with frames of size $4 \times 24$ and $24 \times 4$. Both designs have the same $4 \times 4$ projection grid, hence the same ghost bound~\eqref{eq:crt_ghost}. The non-divisible design, however, enlarges the unaliased region from $4 \times 4$ to $8 \times 8$ bins. It also reduces the PAPR of the point pilot from $13.8$~dB to $10.8$~dB. A point pilot is a train of $N$ pulses over the $MN$ samples of the frame, so its PAPR equals $M$, and the largest delay dimension drops from $24$ to $12$.

\subsection{Non-Swapped and Multi-Frame Designs}

The frames do not necessarily need to be swapped. For example, with $G = 72$, the frames with $(M_1, N_1) = (6, 12)$ and $(M_2, N_2) = (9, 8)$ give $d = 3$ and $L = 18$, and cover $18 \times 24$ bins, i.e., $\rho = 6$, with an unaliased region of $6 \times 8$ bins. Swapped frames cannot cover this spread with $G = 72$, since they require $G = \rho a^2$ for an integer $a$. Dropping the swap constraint therefore enlarges the set of admissible frame sizes, and allows $L$ and $G/d$ to be chosen separately, so that the covered region matches the delay and Doppler spreads of the channel.

The approach also extends to $K > 2$ frames sharing $B$ and $G$. The observations of the same path must then satisfy~\eqref{eq:crt_consistency} for every pair of frames, and the path is recovered with delay index modulo $\mathrm{lcm}(M_1,\dots,M_K)$ and Doppler index modulo $\mathrm{lcm}(N_1,\dots,N_K)$, i.e., the frames cover a spread
\begin{equation}
  \rho = \frac{\mathrm{lcm}(M_1,\dots,M_K)\,\mathrm{lcm}(N_1,\dots,N_K)}{G},
  \label{eq:multi_spread}
\end{equation}
which reduces to~\eqref{eq:crt_spread} for $K = 2$. The unaliased region, of $\min_f M_f \times \min_f N_f$ bins, does not
shrink if each added frame~$f$ has a delay dimension between those of the
frames before it, i.e.,
$\min_{f'<f} M_{f'} \le M_f \le \max_{f'<f} M_{f'}$.
Since a ghost target must now be consistent across all pairs of frames, the
ghost probability can also decrease. The cost is the energy of the
additional pilots.

In the rest of the paper, we implement the proposed design with Zak-OTFS, using the swapped frames of Section~\ref{sec:idea}. Section~\ref{sec:system model} presents the Zak-OTFS system model, and Section~V the dual-frame transmitter and receivers. We then discuss their complexity and compare their performance with the baseline systems, and we finally discuss how spread pilots and data transmission can be supported.

\section{Zak-OTFS System Model}
\label{sec:system model}

Consider a radar scene with $P$ targets, whose DD spreading function is given by
\begin{equation}
\label{sysmod1}
h(\tau,\nu)=\sum_{i=1}^{P}h_i\delta(\tau-\tau_i)\delta(\nu-\nu_i),
\end{equation}
where $h_i$, $\tau_i$, and $\nu_i$ denote the complex gain, delay, and Doppler shift of the $i$-th target, respectively.

The delay and Doppler shift are written as~\cite{sheikh2026variational}\footnote{The integers $k_i = \lfloor B\,\tau_i \rceil$ and
$l_i = \lfloor T\,\nu_i \rceil$ are the delay and Doppler tap indices,
with $\lfloor\cdot\rceil$ denoting rounding to the nearest integer, and the
real values $\tilde k_i, \tilde l_i \in (-0.5, 0.5]$ are the fractional
delay and Doppler parts, respectively.}:
\begin{equation}
\tau_i = \frac{k_i+\tilde k_i}{B}, \qquad
\nu_i = \frac{l_i+\tilde l_i}{T}.
\end{equation}
Let $x(t)$ denote the transmitted radar signal with duration $T$ and bandwidth $B$. Its DD representation is given by the Zak transform $x_{dd}(\tau,\nu)=\mathcal{Z}_t(x(t))
=\sqrt{\tau_p}\sum_{n\in\mathbb{Z}}x(\tau+n\tau_p)e^{-j2\pi n\nu\tau_p}$~\cite{mohammed2022otfs, mohammed2021derivation},
where $\tau_p$ and $\nu_p=1/\tau_p$ denote the delay and Doppler periods, respectively. The discrete DD representation is denoted by $x_{dd}[k,l]$, with delay and Doppler periods $\tau_p=M/B $ and $\nu_p=N/T$, respectively. Lifting this representation to the DD lattice
$\Lambda=\{(k\tau_p/M,l\nu_p/N)\mid k,l\in\mathbb{Z}\}$
gives
\begin{equation}
\label{sysmod4}
\widetilde{x}_{dd}(\tau,\nu)=
\sum_{k,l\in\mathbb{Z}}x_{dd}[k,l]
\delta\left(\tau-\frac{k\tau_p}{M}\right)
\delta\left(\nu-\frac{l\nu_p}{N}\right).
\end{equation}

The continuous DD signal is pulse-shaped to limit the time duration and bandwidth of the transmitted signal, resulting in
\begin{equation}
\label{sysmod5}
\begin{aligned}
x_{dd}(\tau,\nu)
&=w_{tx}(\tau,\nu)*_{\sigma}\widetilde{x}_{dd}(\tau,\nu)\\
&=\iint w_{tx}(\tau',\nu')\widetilde{x}_{dd}(\tau-\tau',\nu-\nu')\\
&\qquad\quad\times e^{j2\pi\nu'(\tau-\tau')}\,d\tau'\,d\nu',
\end{aligned}
\end{equation}
where $w_{tx}(\tau,\nu)$ is the transmit pulse-shaping filter and $*_{\sigma}$ denotes twisted convolution. The transmitted time-domain signal is obtained through the inverse time Zak transform: \begin{equation}
    x(t)=\mathcal{Z}_t^{-1}(x_{dd}(\tau,\nu))
=\sqrt{\tau_p}\int_{0}^{\nu_p}x_{dd}(t,\nu)\,d\nu.
\label{izak}
\end{equation}

The received time-domain signal is given by~\cite{hadani2017otfs}
\begin{equation}
\label{sysmod6}
r(t)=\underbrace{\iint h(\tau,\nu)x(t-\tau)e^{j2\pi\nu(t-\tau)}
\,d\tau\,d\nu}_{r'(t)}+n(t),
\end{equation}
where $n(t)$ is additive white Gaussian noise (AWGN), whose samples at the
receiver have variance $\sigma_w^2$, i.e., $\sigma_w^2$ is the noise power
per received sample.
\begin{figure*}[!t]
  \centering
  \subfloat[]{
\begin{tikzpicture}[
  font=\small,
  >={Latex[length=2.2mm]},
  blk/.style={draw, rounded corners=2pt, align=center, minimum width=2.3cm,
              minimum height=1.25cm, line width=0.6pt, fill=white, inner sep=2pt},
  txblk/.style={blk, draw=green!50!black, fill=green!10},
  rxblk/.style={blk, draw=blue!60!black, fill=blue!8},
  op/.style={draw, circle, line width=0.6pt, inner sep=0pt, minimum size=0.55cm, fill=white},
  sig/.style={->, line width=0.8pt},
  lab/.style={font=\small, inner sep=1.5pt},
]
\newcommand{\ddgrid}[5]{%
  \begin{scope}[shift={(#1,#2)}, x=0.16cm, y=0.16cm]
    \fill[#5!18] ({-#4/2},{-#3/2}) rectangle ({#4/2},{#3/2});
    \draw[step=1, line width=0.25pt, shift={({-#4/2},{-#3/2})}] (0,0) grid (#4,#3);
    \draw[#5, line width=0.8pt] ({-#4/2},{-#3/2}) rectangle ({#4/2},{#3/2});
  \end{scope}}
\def\ya{1.35}
\def\yb{-1.35}

\ddgrid{0.3}{\ya}{9}{3}{frameA}
\ddgrid{0.3}{\yb}{3}{9}{black}
\fill[black] ($(0.3,\ya)+(-0.24,-0.08)$) rectangle ++(0.16,0.16);
\fill[black] ($(0.3,\yb)+(-0.72,-0.08)$) rectangle ++(0.16,0.16);
\node[lab, above=1pt] at (0.3,\ya+0.72) {$x^{(1)}_{dd}[k,l]$};
\node[lab, below=1pt] at (0.3,\yb-0.24) {$x^{(2)}_{dd}[k,l]$};

\node[txblk] (tx1) at (2.95,\ya) {Zak-OTFS\\transmitter\\$(M_1,N_1)$};
\node[txblk] (tx2) at (2.95,\yb) {Zak-OTFS\\transmitter\\$(M_2,N_2)$};
\draw[sig] (0.6,\ya) -- (tx1);
\draw[sig] (1.05,\yb) -- (tx2);

\node[op] (sum) at (5.55,0) {$+$};
\draw[sig] (tx1.east) -- node[lab, above] {$x^{(1)}(t)$} (sum |- 0,\ya) -- (sum.north);
\draw[sig] (tx2.east) -- node[lab, below] {$x^{(2)}(t)$} (sum |- 0,\yb) -- (sum.south);

\node[draw=orange!70!black, fill=orange!22, rounded corners=2pt, line width=0.6pt,
      minimum width=1.6cm, minimum height=1.05cm, align=center] (ch) at (7.2,0)
      {Channel\\$h(\tau,\nu)$};
\draw[sig] (sum.east) -- node[lab, above] {$x(t)$} (ch.west);
\node[op] (noise) at (8.85,0) {$+$};
\draw[sig] (ch.east) -- (noise.west);
\draw[sig] ($(noise)+(0,-0.85)$) node[lab, below] {$n(t)$} -- (noise.south);
\coordinate (split) at (9.6,0);
\draw[line width=0.8pt] (noise.east) -- node[lab, above] {$r(t)$} (split);
\fill (split) circle (1.6pt);

\node[rxblk, minimum width=2.1cm] (rx1) at (10.95,\ya) {Zak-OTFS\\receiver\\$(M_1,N_1)$};
\node[rxblk, minimum width=2.1cm] (rx2) at (10.95,\yb) {Zak-OTFS\\receiver\\$(M_2,N_2)$};
\draw[sig] (split) |- (rx1.west);
\draw[sig] (split) |- (rx2.west);

\ddgrid{13.05}{\ya}{9}{3}{frameA}
\ddgrid{13.05}{\yb}{3}{9}{black}
\node[lab, above=1pt] at (13.05,\ya+0.72) {$y^{(1)}_{dd}[k,l]$};
\node[lab, below=1pt] at (13.05,\yb-0.24) {$y^{(2)}_{dd}[k,l]$};
\draw[sig] (rx1.east) -- (12.75,\ya);
\draw[sig] (rx2.east) -- (12.31,\yb);

\node[draw=blue!75!black, fill=blue!5, rounded corners=2pt, line width=0.7pt,
      minimum width=1.8cm, minimum height=3.9cm, align=center] (det)
      at (15.0,0) {Overspread\\target\\detection};
\draw[sig] (13.35,\ya) -- (det.west |- 0,\ya);
\draw[sig] (13.79,\yb) -- (det.west |- 0,\yb);
\draw[sig] (det.east) -- ++(0.35,0) node[right, align=left, inner sep=1pt] {$(\hat\tau_i,\hat\nu_i,\hat h_i)$};
\end{tikzpicture}\label{fig:system_overview}}

  \vspace{3mm}

  \subfloat[]{
\begin{tikzpicture}[
  font=\small,
  >={Latex[length=2.2mm]},
  blk/.style={draw, rounded corners=2pt, align=center, minimum width=1.3cm,
              minimum height=1.0cm, line width=0.6pt, fill=white, inner sep=2pt},
  txblk/.style={blk, draw=green!50!black, fill=green!10},
  rxblk/.style={blk, draw=blue!60!black, fill=blue!8},
  op/.style={draw, circle, line width=0.6pt, inner sep=0pt, minimum size=0.55cm, fill=white},
  sig/.style={->, line width=0.8pt},
  lab/.style={font=\small, inner sep=1.5pt},
  slab/.style={font=\footnotesize, inner sep=1pt},
]
\node[lab, anchor=east] (xin) at (0.9,0) {$x^{(f)}_{dd}[k,l]$};
\node[op] (lift) at (1.55,0) {$\times$};
\node[txblk] (wtx) at (3.95,0) {Twisted\\conv.};
\node[txblk] (izt) at (6.55,0) {Inverse\\Zak};
\draw[sig] (xin.east) -- (lift);
\draw[sig] (lift) -- node[slab, above] {$\tilde{x}^{(f)}_{dd}(\tau,\nu)$} (wtx);
\draw[sig] (wtx) -- node[slab, above] {$x^{(f)}_{dd}(\tau,\nu)$} (izt);
\draw[sig] ($(lift)+(0,0.8)$) node[lab, above, align=center]
      {impulse train\\$(\tau_{pf},\nu_{pf})$} -- (lift);
\draw[sig] ($(wtx)+(0,1.0)$) node[lab, above] {$w_{tx}(\tau,\nu)$} -- (wtx);
\draw[sig] (izt.east) -- node[lab, above] {$x^{(f)}(t)$} ++(0.95,0);

\node[lab, anchor=west] (rin) at (8.55,0) {$r(t)$};
\node[rxblk, minimum width=1.1cm] (zt) at (10.25,0) {Zak};
\node[rxblk] (wrx) at (12.75,0) {Twisted\\conv.};
\node[rxblk] (smp) at (15.35,0) {DD\\sampling};
\draw[sig] (rin.east) -- (zt);
\draw[sig] (zt) -- node[slab, above] {$r^{(f)}_{dd}(\tau,\nu)$} (wrx);
\draw[sig] (wrx) -- node[slab, above] {$y^{(f)}_{dd}(\tau,\nu)$} (smp);
\draw[sig] ($(wrx)+(0,1.0)$) node[lab, above] {$w_{rx}(\tau,\nu)$} -- (wrx);
\node[lab, below] at (zt.south) {$(M_f,N_f)$};
\draw[sig] (smp.east) -- ++(0.35,0) node[lab, right, inner sep=1pt] {$y^{(f)}_{dd}[k,l]$};

\begin{scope}[on background layer]
  \draw[black!40, line width=0.5pt, dashed, rounded corners=5pt] (-0.35,-0.95) rectangle (8.25,1.8);
  \node[black!70, anchor=north west] at (-0.3,-0.97) {Zak-OTFS transmitter};
  \draw[black!40, line width=0.5pt, dashed, rounded corners=5pt] (8.45,-0.95) rectangle (17.8,1.8);
  \node[black!70, anchor=north east] at (17.75,-0.97) {Zak-OTFS receiver};
\end{scope}
\end{tikzpicture}\label{fig:zak_blocks}}
\caption{Superimposed dual-frame Zak-OTFS sensing. (a)~The two frames, of sizes $(M_1,N_1)$ and $(M_2,N_2)$, are generated by two Zak-OTFS transmitters and summed into $x(t)$. The received signal $r(t)$ is processed by two Zak-OTFS receivers, and the two DD grids are combined for overspread target detection. (b)~Zak-OTFS transmitter and receiver of frame~$f$.}
  \label{fig:system}
\end{figure*}
The receiver applies the time Zak transform,
\begin{equation}
\begin{aligned}
      r_{dd}(\tau,\nu) = \mathcal{Z}_t\big(r(t)\big)&
  = \sqrt{\tau_p}\sum_{n\in\mathbb{Z}} r(\tau + n\tau_p)\,
  e^{-j2\pi n\nu\tau_p}\\
  &= r'_{dd}(\tau,\nu)+\mathcal{Z}_t\big(n(t)\big),
\end{aligned}
  \label{zak}
\end{equation}
where $r'_{dd}(\tau,\nu)=\mathcal{Z}_t(r'(t))$, and applies the matched
filter
$w_{rx}(\tau,\nu)\triangleq w_{tx}^{*}(-\tau,-\nu)e^{j2\pi\nu\tau}$, yielding
\begin{equation}
\label{sysmod7}
\begin{aligned}
    y_{dd}(\tau,\nu)
&=w_{rx}(\tau,\nu)*_{\sigma}r_{dd}(\tau,\nu)\\&=w_{rx}(\tau,\nu)*_{\sigma}r'_{dd}(\tau,\nu)+n_{dd}(\tau,\nu),
\end{aligned}
\end{equation}
where
$n_{dd}(\tau,\nu)=w_{rx}(\tau,\nu)*_{\sigma}\mathcal{Z}_t(n(t))$
is the colored noise.
Sampling on $\Lambda$ gives
\begin{equation}
\label{sysmod8}
y_{dd}[k,l]=
y_{dd}\left(\frac{k\tau_p}{M},\frac{l\nu_p}{N}\right).
\end{equation}

Using equations \eqref{sysmod1} to \eqref{sysmod8}, the discrete DD-domain input--output (I/O) relation is given by \cite{mohammed2022otfs, mohammed2023predictability, mohammed2024otfsbook,dan2026zakotfs}
\begin{equation}
\label{sysmod9}
\begin{aligned}
y_{dd}[k,l]
&=h[k,l]*_{\sigma}x_{dd}[k,l]+n_{dd}[k,l]\\
&=\sum_{k',l'\in\mathbb{Z}}h[k',l']x_{dd}[k-k',l-l']\\
&\quad\times e^{j2\pi\frac{l'(k-k')}{MN}}+n_{dd}[k,l],
\end{aligned}
\end{equation}
where
$n_{dd}[k,l]\triangleq n_{dd}\left(\frac{k\tau_p}{M},\frac{l\nu_p}{N}\right)$,
$h[k,l]\triangleq h_{dd}\left(\frac{k\tau_p}{M},\frac{l\nu_p}{N}\right)$,
and $h_{dd}(\tau,\nu)\triangleq
w_{rx}(\tau,\nu)*_{\sigma}h(\tau,\nu)*_{\sigma}w_{tx}(\tau,\nu).$
Thus, $h(\tau,\nu)$ describes the target locations in the continuous DD domain, whereas $h[k,l]$ represents the effective channel response on the discrete DD grid. The peaks of $|h[k,l]|$ indicate the target locations.

The transmit filter $w_{\mathrm{tx}}(\tau,\nu)$ limits the
signal duration and bandwidth.
We use the unit-energy Gauss--sinc filter~\cite{ArpanDas26}, whose parameters
confine $99\%$ of the transmitted signal energy within duration
$T$ and bandwidth $B$.
From \eqref{sysmod1} and \eqref{sysmod9}, the effective channel can be expressed as $
    h[k,l]=\sum_{i=1}^{P}h_i[k,l]
$,
where $h_i[k,l]\triangleq h_{dd,i}\left(\frac{k\tau_p}{M},\frac{l\nu_p}{N}\right)$ and
$h_{dd,i}(\tau,\nu)\triangleq h_i\,w_{rx}(\tau,\nu)
*_{\sigma}\big[\delta(\tau-\tau_i)\delta(\nu-\nu_i)\big]
*_{\sigma}w_{tx}(\tau,\nu).$
The SNR of the received signal reflected from the $i$-th target is therefore given by 
\begin{equation}
\gamma_i\triangleq
\frac{\sum_{k=0}^{M-1}\sum_{l=0}^{N-1}|x_{dd,h_i}[k,l]|^2}
{\sum_{k=0}^{M-1}\sum_{l=0}^{N-1}\mathbb{E}[|n_{dd}[k,l]|^2]}\ , \end{equation}
where $x_{dd,h_i}[k,l]\triangleq h_i[k,l]*_{\sigma}x_{dd}[k,l]$.

\section{Proposed Dual-Frame Zak-OTFS Sensing Receivers}
\label{sec:proposed_detection}

We now implement the frame structure of Section~\ref{sec:idea} with Zak-OTFS. The two frames can be transmitted in consecutive time intervals, on different frequency bands, or superimposed in the same time--frequency resource. We consider the last option, which requires no additional time or bandwidth, and whose architecture is shown in Fig.~\ref{fig:system}. Superimposing the two frames, which use Zak-OTFS bases, however, introduces interference between their pilots, and the main task of the receiver is to detect the targets in the presence of this interference and to pair their detections in the two frames.

\subsection{Transmitter and Receiver}

frame~1 has size $(M_1, N_1) = (a, b)$ and frame~2 has size $(M_2, N_2) = (b, a)$, with $\rho = b/a$ an integer and $M_fN_f = G = ab,\ f\in\{1,2\}$, as in Section~\ref{sec:idea}. Each frame carries a point pilot in its DD grid. As shown in Fig.~\ref{fig:zak_blocks}, the DD signal $x^{(f)}_{dd}[k,l]$ of frame~$f$ is lifted to the DD lattice of the frame, with delay period $\tau_{pf} = M_f/B$, as in~\eqref{sysmod4}, filtered by the transmit filter $w_{tx}(\tau,\nu)$ as in~\eqref{sysmod5}, and converted to the time domain by the inverse Zak transform~\eqref{izak}, which gives the transmit signal $x^{(f)}(t)$. Both signals have bandwidth $B$ and duration $T = G/B$, measured as the support that contains $99\%$ of the energy of the Gauss--sinc filter, and they are summed into the transmit signal $x(t)$. In the time domain, each pilot is a pulse train modulated by a tone, i.e., a pulsone~\cite{mohammed2022otfs}: the pilot of frame~1 consists of $b$ pulses spaced $\tau_{p1} = a/B$ apart, and the pilot of frame~2 of $a$ pulses spaced $\tau_{p2} = b/B$ apart, with each pulse shaped by the transmit filter. The frequency of the tone is set by the Doppler position of the pilot, and is zero for a pilot at zero Doppler.

At the receiver, the received signal $r(t)$ is processed by two Zak-OTFS receivers of sizes $(M_1,N_1)$ and $(M_2,N_2)$. Each receiver applies the Zak transform~\eqref{zak}, the receive filter $w_{rx}(\tau,\nu)$ as in~\eqref{sysmod7}, and sampling on its own DD lattice as in~\eqref{sysmod8}, which gives the received DD signals $y^{(1)}_{dd}[k,l]$ and $y^{(2)}_{dd}[k,l]$.

\subsection{Cross-Pilot Interference}

Since the two frames are superimposed, each DD frame grid also receives the pilot of the other frame. The pilot of frame~2 is a pulse train with one pulse every $b = \rho a$ samples, i.e., one pulse every $\rho$ delay periods of frame~1. In the frame-1 grid, a path at delay bin $k$ and Doppler bin $l$ therefore produces, in addition to its own peak, a comb of $\rho$ peaks at $\big(k \bmod a,\ (l+ja) \bmod b\big)$, $j = 0,\dots,\rho-1$, each with amplitude $1/\sqrt{\rho}$ relative to the own peak. The comb peak $j = 0$ falls on the own peak and, since the two pilots are placed at the same DD position with the same phase, adds coherently to it, giving a combined amplitude $1+1/\sqrt{\rho}$. The same holds in the frame-2 grid, where the comb lies along the delay axis. Fig.~\ref{fig:comb} shows the own peak and the comb of a single path in the frame-1 grid for $\rho = 4$ and $\rho = 16$.

The comb peaks have the same magnitude pattern for every path, but their phases cannot be observed before the path is estimated. The comb of a path can therefore be removed only after its delay, Doppler, and gain have been estimated, which is the approach of the DS-XC receiver described below. If the two frames are instead transmitted on orthogonal resources, e.g., on separate frequency channels, in consecutive frames, or with orthogonal spreading sequences, no cross-pilot interference arises.
\begin{figure}[!t]
  \centering
  \resizebox{1\columnwidth}{!}{
\begin{tikzpicture}[font=\footnotesize]
\begin{scope}[yshift=0.0cm]
\fill[black!100] (0.6900,1.1000) rectangle (0.9200,1.3200);
\fill[black!33] (2.5300,1.1000) rectangle (2.7600,1.3200);
\fill[black!33] (4.3700,1.1000) rectangle (4.6000,1.3200);
\fill[black!33] (6.2100,1.1000) rectangle (6.4400,1.3200);
\draw[black!25,step=1] (0,0) rectangle (7.3600,1.7600);
\draw[black!12] (0,0.2200) -- (7.3600,0.2200);
\draw[black!12] (0,0.4400) -- (7.3600,0.4400);
\draw[black!12] (0,0.6600) -- (7.3600,0.6600);
\draw[black!12] (0,0.8800) -- (7.3600,0.8800);
\draw[black!12] (0,1.1000) -- (7.3600,1.1000);
\draw[black!12] (0,1.3200) -- (7.3600,1.3200);
\draw[black!12] (0,1.5400) -- (7.3600,1.5400);
\draw[frameA,thick] (0.8050,1.2100) circle (0.13);
\draw[blue!70!black] (2.6450,1.2100) circle (0.13);
\draw[blue!70!black] (4.4850,1.2100) circle (0.13);
\draw[blue!70!black] (6.3250,1.2100) circle (0.13);
\draw[{Stealth[length=3pt]}-{Stealth[length=3pt]}] (0.8050,1.4300) -- (2.6450,1.4300) node[midway,above,inner sep=1pt] {$a$};
\node[anchor=south west,inner sep=1pt] at (0,1.8100) {(a) $a\times b = 8\times32$, $\rho = 4$: own peak and $3$ comb peaks};
\node[rotate=90,anchor=south] at (-0.08,0.8800) {Delay};
\end{scope}
\begin{scope}[yshift=-1.63cm]
\fill[black!100] (0.3450,0.2200) rectangle (0.4600,0.4400);
\fill[black!20] (0.8050,0.2200) rectangle (0.9200,0.4400);
\fill[black!20] (1.2650,0.2200) rectangle (1.3800,0.4400);
\fill[black!20] (1.7250,0.2200) rectangle (1.8400,0.4400);
\fill[black!20] (2.1850,0.2200) rectangle (2.3000,0.4400);
\fill[black!20] (2.6450,0.2200) rectangle (2.7600,0.4400);
\fill[black!20] (3.1050,0.2200) rectangle (3.2200,0.4400);
\fill[black!20] (3.5650,0.2200) rectangle (3.6800,0.4400);
\fill[black!20] (4.0250,0.2200) rectangle (4.1400,0.4400);
\fill[black!20] (4.4850,0.2200) rectangle (4.6000,0.4400);
\fill[black!20] (4.9450,0.2200) rectangle (5.0600,0.4400);
\fill[black!20] (5.4050,0.2200) rectangle (5.5200,0.4400);
\fill[black!20] (5.8650,0.2200) rectangle (5.9800,0.4400);
\fill[black!20] (6.3250,0.2200) rectangle (6.4400,0.4400);
\fill[black!20] (6.7850,0.2200) rectangle (6.9000,0.4400);
\fill[black!20] (7.2450,0.2200) rectangle (7.3600,0.4400);
\draw[black!25,step=1] (0,0) rectangle (7.3600,0.8800);
\draw[black!12] (0,0.2200) -- (7.3600,0.2200);
\draw[black!12] (0,0.4400) -- (7.3600,0.4400);
\draw[black!12] (0,0.6600) -- (7.3600,0.6600);
\draw[frameA,thick] (0.4025,0.3300) circle (0.13);
\draw[blue!70!black] (0.8625,0.3300) circle (0.13);
\draw[blue!70!black] (1.3225,0.3300) circle (0.13);
\draw[blue!70!black] (1.7825,0.3300) circle (0.13);
\draw[blue!70!black] (2.2425,0.3300) circle (0.13);
\draw[blue!70!black] (2.7025,0.3300) circle (0.13);
\draw[blue!70!black] (3.1625,0.3300) circle (0.13);
\draw[blue!70!black] (3.6225,0.3300) circle (0.13);
\draw[blue!70!black] (4.0825,0.3300) circle (0.13);
\draw[blue!70!black] (4.5425,0.3300) circle (0.13);
\draw[blue!70!black] (5.0025,0.3300) circle (0.13);
\draw[blue!70!black] (5.4625,0.3300) circle (0.13);
\draw[blue!70!black] (5.9225,0.3300) circle (0.13);
\draw[blue!70!black] (6.3825,0.3300) circle (0.13);
\draw[blue!70!black] (6.8425,0.3300) circle (0.13);
\draw[blue!70!black] (7.3025,0.3300) circle (0.13);
\draw[{Stealth[length=3pt]}-{Stealth[length=3pt]}] (0.4025,0.5500) -- (0.8625,0.5500) node[midway,above,inner sep=1pt] {$a$};
\node[anchor=south west,inner sep=1pt] at (0,0.9300) {(b) $a\times b = 4\times64$, $\rho = 16$: own peak and $15$ comb peaks};
\node[rotate=90,anchor=south] at (-0.08,0.4400) {Delay};
\end{scope}
\node[anchor=north] at (3.6800,-1.6500) {Doppler};
\begin{scope}[yshift=-2.1799999999999997cm]
\draw[frameA,thick] (0.4,0) circle (0.13); \node[anchor=west] at (0.6,0) {Own peak (pilot of frame 1)};
\draw[blue!70!black] (4.75,0) circle (0.13); \node[anchor=west] at (4.95,0) {Comb (pilot of frame 2)};
\end{scope}
\end{tikzpicture}}
\caption{Magnitude of $y^{(1)}_{dd}[k,l]$ for a single path with superimposed pilots, $G = 256$: (a)~$\rho = 4$ and (b)~$\rho = 16$. The pilot of frame~2 appears as a comb of $\rho - 1$ peaks spaced by $a$ Doppler bins, each with a fraction $1/\rho$ of its power.}
  \label{fig:comb}
\end{figure}

\subsection{Pairing the Detections of the Two Frames}

The detections of the two frames are paired in three steps. First, each detection is projected onto the overlap square by reducing its delay and Doppler indices modulo $a$, as in~\eqref{eq:consistency}. Two detections of different frames are consistent if their projections lie within half a bin of each other, and the connected components of this consistency relation form groups of detections that may belong to the same targets.

Second, within each group, the detections are paired by their gains. Let $\hat h^{(1)}_i$ and $\hat h^{(2)}_j$ be the gains of the $i$-th detection of frame~1 and of the $j$-th detection of frame~2. Since the two detections of the same target are scaled by the same path gain, the cost of pairing them is their relative amplitude difference,
\begin{equation}
  c_{ij} = \frac{\big|\,|\hat h^{(1)}_i| - |\hat h^{(2)}_j|\,\big|}
               {\max\big(|\hat h^{(1)}_i|, |\hat h^{(2)}_j|\big)} \in [0,1],
  \label{eq:gain_cost}
\end{equation}
with $c_{ij} = \infty$ for inconsistent pairs. The Hungarian algorithm~\cite{munkres1957assignment} then finds the one-to-one assignment $\mathcal{A}$ that minimizes the total cost $J(\mathcal{A}) = \sum_{(i,j) \in \mathcal{A}} c_{ij}$.

Third, since the Hungarian algorithm always returns a single assignment, even when another one is almost as good, the reliability of each assignment is tested. Swapping the partners of two assigned pairs $(i,j)$ and $(i',j')$ changes the total cost by
\begin{equation}
  \Delta = c_{ij'} + c_{i'j} - c_{ij} - c_{i'j'} \ge 0,
  \label{eq:swap_cost}
\end{equation}
where the inequality follows the optimality of $\mathcal{A}$. If $\Delta \ge \theta$ for every swap, the assignment is retained as unambiguous. If $\Delta < \theta$, the two targets have similar gains, so the assignment cannot be decided: both pairings are kept, which gives two targets and two ghosts, and the four candidates are flagged as an ambiguous set. The threshold is $\theta = \theta_g$ for the final output and $\theta = \theta_{xc}$ inside the iterations of the DS-XC receiver. Finally, each paired target is placed at the delay measured by frame~2 and the Doppler measured by frame~1, as in~\eqref{eq:reconstruction}, with gain $(\hat h^{(1)}_i + \hat h^{(2)}_j)/2$.

\subsection{Peak Detection and Off-Grid Refinement for Fractional Delay and Doppler Shifts}
In a practical channel, the delays and Doppler shifts of the paths are
fractional, i.e., they do not fall on the DD grid, and the energy of each
path spreads over the neighboring bins. In this subsection, we describe how the
paths are detected and how their fractional delays and Doppler shifts are
estimated by refining the position of each detected peak separately in delay
and Doppler, as illustrated in Fig.~\ref{fig:frac}.
\begin{figure}[!t]
  \centering
  \resizebox{0.8\columnwidth}{!}{\begin{tikzpicture}[scale=1.6,font=\LARGE]
  \def\dx{0.42}\def\dy{0.36}
  \foreach \i in {-2,...,2}{
    \foreach \j in {-2,...,2}{
      \pgfmathsetmacro{\ax}{0.55*abs(\i-\dx)}
      \pgfmathsetmacro{\ay}{0.55*abs(\j-\dy)}
      \pgfmathsetmacro{\rx}{\ax<0.001 ? 1 : abs(sin(180*\ax)/(pi*\ax))}
      \pgfmathsetmacro{\ry}{\ay<0.001 ? 1 : abs(sin(180*\ay)/(pi*\ay))}
      \pgfmathsetmacro{\v}{75*\rx*\ry}
      \fill[blue!\v!white] (\i-0.5,\j-0.5) rectangle (\i+0.5,\j+0.5); \draw[gray!50] (\i-0.5,\j-0.5) rectangle (\i+0.5,\j+0.5);
    }}
  
  \draw[line width=2pt,red] (-0.5,-0.5) rectangle (0.5,0.5);
  \draw[line width=2pt,red,dashed] (0.5,-0.5) rectangle (1.5,0.5);
  \draw[line width=2pt,red,dotted] (-0.5,0.5) rectangle (0.5,1.5);
  \fill (0,0) circle (1.2pt);
  \draw[line width=1.6pt] (\dx-0.13,\dy-0.13) -- (\dx+0.13,\dy+0.13) (\dx-0.13,\dy+0.13) -- (\dx+0.13,\dy-0.13);
  \draw[|-|,line width=1.2pt] (0,-0.9) -- (\dx,-0.9) node[midway,below=1pt,font=\Huge] {$\delta$};
  \draw[densely dotted] (\dx,-0.95) -- (\dx,\dy);
  \draw[|-|,line width=1.2pt] (-0.9,0) -- (-0.9,\dy) node[midway,left=1pt,font=\Huge] {$\epsilon$};
  \draw[densely dotted] (-0.95,\dy) -- (\dx,\dy);


  \node[below,font=\Huge] at (0,-2.8) {$k$};
  \node[below,font=\Huge] at (1,-2.8) {$k{+}1$};
  \node[left,font=\Huge] at (-2.8,0) {$l$};
  \node[left,font=\Huge] at (-2.8,1) {$l{+}1$};
  \foreach \t in {0,1}{\draw (\t,-2.75) -- (\t,-2.85); \draw (-2.75,\t) -- (-2.85,\t);}
  \draw[->] (-2.5,-2.8) -- (2.8,-2.8) node[right,font=\Huge] {Delay};
  \draw[->] (-2.8,-2.5) -- (-2.8,2.8) node[above,font=\Huge] {Doppler};
  \begin{scope}[shift={(2.8,2.1)},font=\huge]
    \draw[line width=2pt,red] (0,0) rectangle (0.3,0.3); \node[right] at (0.4,0.15) {Peak};
    \draw[line width=2pt,red,dashed] (0,-0.6) rectangle (0.3,-0.3); \node[right] at (0.4,-0.45) {Delay neighbor};
    \draw[line width=2pt,red,dotted] (0,-1.2) rectangle (0.3,-0.9); \node[right] at (0.4,-1.05) {Doppler neighbor};
    \draw[line width=1.6pt] (0.02,-1.88) -- (0.28,-1.62) (0.02,-1.62) -- (0.28,-1.88); \node[right] at (0.4,-1.75) {Target};
  \end{scope}
\end{tikzpicture}}
\caption{Off-grid refinement: the target position is estimated from the
  peak bin and its larger neighbors, separately for the fractional delay
  $\delta$ and the fractional Doppler $\epsilon$.}
  \label{fig:frac}
\end{figure}
In the DD grid of frame~$f$, the targets are detected as the local maxima of the normalized power $|y^{(f)}_{dd}[k,l]|^2/\sigma_w^2$, for $k = 0, \dots, M_f - 1$ and $l = 0, \dots, N_f - 1$, that exceed the constant-false-alarm-rate threshold $T_{\mathrm{det}} = \ln(G/P_{\mathrm{FA}})$. Under noise only, $|y^{(f)}_{dd}[k,l]|^2/\sigma_w^2$ is exponentially distributed with unit mean, so each of the $G$ bins exceeds $T_{\mathrm{det}}$ with probability $e^{-T_{\mathrm{det}}} = P_{\mathrm{FA}}/G$. The design parameter $P_{\mathrm{FA}}$ is therefore the expected number of FA per grid due to noise alone. To suppress the sidelobes of off-grid paths, a peak is discarded if a peak within two bins exceeds it by more than $15$~dB.

A path generally lies between bins, so the position of each detected peak is refined separately in delay and Doppler. Let the peak lie at bin $(k,l)$, and let the larger of its two delay neighbors be at $(k+1,l)$, so that the target delay is $(k+\delta)/B$ for some $\delta \in [0, 1/2]$. The transmit and receive filters are separable in delay and Doppler (Section~\ref{sec:system model}), and we let $m(x)$ denote the magnitude of the delay response of their cascade, with $x$ in bins. The ratio
\begin{equation}
  \eta = \frac{|y^{(f)}_{dd}[k+1,l]|}{|y^{(f)}_{dd}[k,l]| + |y^{(f)}_{dd}[k+1,l]|}
  \approx \frac{m(1-\delta)}{m(\delta) + m(1-\delta)}
  \label{eq:refine}
\end{equation}
increases monotonically with $\delta$, so the measured $\eta$ is mapped to $\delta$, and the refined delay is $(k+\delta)/B$. If the larger neighbor is at $(k-1,l)$, the same rule gives the delay $(k-\delta)/B$. The Doppler shift is refined in the same way along the Doppler axis, using the Doppler response of the filters. The value at the peak bin provides the gain estimate used for pairing.
\begin{figure}[!htbp]
  \centering
  \begin{tikzpicture}[
  node distance=4mm,
  every node/.style={font=\footnotesize},
  base/.style={rounded corners=2pt, align=center, line width=0.7pt,
               execute at begin node={\hyphenpenalty=10000\relax},
               minimum height=6mm, inner sep=2.2pt},
  blk/.style={base, text width=#1, draw=blue!55!black, fill=blue!5},
  blk/.default=42mm,
  acc/.style={base, text width=#1, draw=green!45!black, fill=green!8},
  acc/.default=42mm,
  dec/.style={draw=orange!75!black, fill=orange!10, line width=0.7pt, diamond,
              aspect=4, align=center, inner sep=0.3pt},
  term/.style={draw=black!70, fill=black!7, line width=1.4pt, rounded corners=3mm,
               align=center, minimum width=24mm, minimum height=5.5mm},
  arr/.style={-{Stealth[length=2mm]}, line width=0.6pt, black!75},
  lbl/.style={font=\footnotesize\itshape, inner sep=1pt, black!75},
  ttl/.style={font=\small\bfseries},
  stp/.style={font=\footnotesize\bfseries, text=black, anchor=east, inner sep=1pt},
  frm/.style={draw=#1, dashed, line width=1.1pt, rounded corners=3pt,
              dash pattern=on 4pt off 2pt},
  flbl/.style={font=\small\bfseries, text=#1, inner sep=1pt, anchor=south, rotate=90},
]
\colorlet{loopc}{blue!60!black}
\colorlet{ghostc}{red!60!black}
\coordinate (XL) at (-25mm,0);
\coordinate (XR) at ( 25mm,0);

\node[ttl]  (ta)                   {(a) DS};
\node[term] (a0) [below=2mm of ta] {Grids $y^{(1)}_{dd}, y^{(2)}_{dd}$};
\node[blk]  (a1) [below=of a0]     {Detection (threshold $T_{\mathrm{det}}$)\\+ Off-grid refinement};
\node[blk]  (a2) [below=of a1]     {Grouping by projection onto the overlap square};
\node[blk]  (a3) [below=of a2]     {Hungarian matching and ambiguity test ($\theta_g$ threshold)};
\node[term] (a5) [below=of a3]     {Targets $(\hat\tau^{(2)}_i, \hat\nu^{(1)}_i, \hat h_i)$};
\foreach \u/\v in {a0/a1,a1/a2,a2/a3,a3/a5} \draw[arr] (\u) -- (\v);

\node[ttl]  (tb)    [below=7mm of a5]    {(b) DS-XC};
\node[term] (b0)    [below=2mm of tb]    {Grids $y^{(1)}_{dd}, y^{(2)}_{dd}$};
\node[blk]  (canc)  [below=6mm of b0]    {Cancellation of the combs of accepted targets};
\node[blk]  (det)   [below=of canc]      {DS (a), $\theta_{xc}$ threshold};
\node[blk]  (gain)  [below=of det]       {Gain update (joint LS)};
\node[dec]  (unamb) [below=of gain]      {New unambiguous pair?};
\node[acc]  (accu)  [below=of unamb]     {Accept strongest pair per bin (provisional gain)};
\node[dec]  (conv)  [below=of accu]      {Gains converged?};
\node[dec]  (amb)   [below=of conv]      {New ambiguous sets?};
\node[acc]  (acca)  [below=of amb]       {Accept all candidates and gain update with joint LS};
\node[blk]  (fin)   [below=12mm of acca] {DS (a), $\theta_{g}$ threshold};
\node[blk]  (ls)    [below=6mm of fin]   {Joint LS over all candidates};
\node[blk]  (gho)   [below=of ls]        {In each ambiguous set, discard as ghosts the candidates with low LS gain ($|\hat h| < 0.3\max$)};
\node[term] (bout)  [below=6mm of gho]   {Targets $(\hat\tau^{(2)}_i, \hat\nu^{(1)}_i, \hat h_i)$};

\draw[arr] (b0) -- (canc);
\draw[arr] (canc) -- (det);
\draw[arr] (det) -- (gain);
\draw[arr] (gain) -- (unamb);
\draw[arr] (unamb) -- node[lbl, right, pos=0.4] {yes} (accu);
\draw[arr] (conv) -- node[lbl, right, pos=0.4] {yes} (amb);
\draw[arr] (amb) -- node[lbl, right, pos=0.4] {yes} (acca);
\foreach \u/\v in {fin/ls,ls/gho,gho/bout} \draw[arr] (\u) -- (\v);

\draw[arr] (unamb.west) -- node[lbl, above] {no} (XL |- unamb.west) -- (XL |- conv.west) -- (conv.west);
\draw[arr] (accu.east) -- (XR |- accu.east);
\draw[arr] (conv.east) -- node[lbl, above] {no} (XR |- conv.east);
\draw[arr] (acca.east) -- (XR |- acca.east) -- (XR |- canc.east) -- (canc.east);
\coordinate (ex) at ($(acca.south)+(0,-8mm)$);
\draw[arr] (amb.west) -- node[lbl, above] {no} (XL |- amb.west) -- (XL |- ex) -- (ex)
  -- node[lbl, right, pos=0.2, yshift=1mm] {updated grids $\tilde y^{(1)}_{dd}, \tilde y^{(2)}_{dd}$} (fin.north);

\draw[frm=loopc] ($(XL |- canc.north)+(-3mm,2.5mm)$) rectangle ($(XR |- acca.south)+(3mm,-2.5mm)$);
\node[flbl=loopc] at ($(XL |- canc.north)!0.5!(XL |- acca.south)+(-3.8mm,0)$) {Cancellation loop};
\draw[frm=ghostc] ($(XL |- ls.north)+(-3mm,2.5mm)$) rectangle ($(XR |- gho.south)+(3mm,-2.5mm)$);
\node[flbl=ghostc] at ($(XL |- ls.north)!0.5!(XL |- gho.south)+(-3.8mm,0)$) {Ghost removal};

\coordinate (XS) at (-33mm,0);
\node[stp] at (XS |- canc) {Step~1};
\node[stp] at (XS |- det)  {Step~2};
\node[stp] at (XS |- gain) {Step~3};
\node[stp] at (XS |- accu) {Step~4};
\node[stp] at (XS |- acca) {Step~5};
\end{tikzpicture}
  \caption{Flowcharts of the dual-frame receivers: (a)~DS and (b)~DS-XC.}
  \label{fig:flow_receivers}
\end{figure}

\subsection{Receiver Variants}

We consider two receivers, which differ in how they handle the cross-pilot interference and the ambiguous sets. Their flowcharts are shown in Fig.~\ref{fig:flow_receivers}.

The dual-frame sensing (DS) receiver performs the detection and pairing described above once, with the comb peaks left in the grids. Since the own peak exceeds each comb peak by a factor $1+\sqrt{\rho}$, the true pairs are still matched correctly. The comb peaks, however, are themselves detected, and since they have similar gains, their pairings are flagged as ambiguous and all of them are reported. DS therefore has low complexity but produces many FA when $\rho$ is large.

The DS with cross-pilot cancellation (DS-XC) receiver iteratively cancels the combs of the detected targets from the grids, and then removes the ghost targets of the remaining ambiguous sets. For each accepted target~$c$, we define the template $\mathbf{t}_c$ as the received signal that a unit-gain path at $(\hat\tau_c, \hat\nu_c)$ would produce from the transmit signal $x(t)$. The comb of target~$c$ in each grid is the cross-pilot part of $\mathbf{t}_c$ in that grid, scaled by its gain $\hat h_c$. Since the combs are predicted at the estimated positions, DS-XC relies on the accurate sub-bin estimates provided by~\eqref{eq:refine}.

The cancellation loop starts from the received grids with no accepted target, and each iteration performs the following steps:
\begin{enumerate}[label=\textbf{Step~\arabic*:}, leftmargin=*, align=left]
\item The combs of all accepted targets, scaled by their current gains, are subtracted from the received grids $y^{(1)}_{dd}[k,l]$ and $y^{(2)}_{dd}[k,l]$, while their own peaks are kept.
\item DS is applied to the comb-canceled grids, with ambiguity threshold $\theta_{xc}$.
  \item The gains of all accepted targets are re-estimated jointly by least squares (LS) as $\hat{\mathbf{h}} = \mathbf{T}^{+}\mathbf{r}$, where $\mathbf{T} = [\mathbf{t}_1, \dots, \mathbf{t}_C]$ collects the templates of the $C\le \hat{P}$ accepted targets, and $\mathbf{r}$ collects the received samples. The gains read from the peaks are not accurate enough to cancel the combs, whereas the joint fit, in which each template models the full response of a path, separates the contributions of different targets.
  \item In each projection bin, the strongest unambiguous pair that has not yet been accepted, if any, is accepted. Its provisional gain is the pairing gain divided by $1+1/\sqrt{\rho}$, which removes the contribution of its own comb to its own peak.
  \item Only when no unambiguous pair is left and the gains have converged, the remaining ambiguous sets are accepted, so that no detection is missed. All four candidates of each set are accepted and their gains are fitted jointly by LS. Since the ghosts correspond to no path of the channel, the fit assigns them gains close to zero, and their combs are not canceled.
\end{enumerate}
The loop returns to \textbf{Step~1} whenever a target has been accepted or a gain has changed. Otherwise, it stops, and DS is applied a last time to the comb-canceled grids $\tilde y^{(1)}_{dd}[k,l]$ and $\tilde y^{(2)}_{dd}[k,l]$.
The ambiguous sets of this final pass still contain two true targets and two ghosts each. In the ghost removal stage, a joint LS fit over all reported candidates estimates their gains, and, within each ambiguous set, the candidates whose gain magnitude is below $0.3$ times the largest in the set are discarded as ghosts. A set whose remaining candidates no longer share a detection forms a consistent assignment, and is reported as unambiguous.

\section{Computational Complexity}
\label{sec:complexity}
 The complexities are summarized in Table~\ref{tab:complexity}.
In conventional radar architectures, the delay--Doppler response is
obtained by cross-correlating the received signal with the transmitted
waveform, whose computational complexity is quadratic in the
time--bandwidth product, $\mathcal{O}(B^2T^2)$. This holds both for
continuous radar with LFM or FMCW
waveforms~\cite{auslander1985radar,moran2001group,jankiraman2018fmcw}, and
for discrete radar with phase-coded
waveforms~\cite{benedetto2009phase,bai2024design}, since in neither case
the cross-ambiguity has a structure that reduces its
computation~\cite{NishantDiscreteRadar25}. When the transmit waveform is
the Zak-OTFS pulsone, or one of its symplectic transformations, the
cross-ambiguity reduces to $M$ FFTs of length $N$, and its complexity is
$\mathcal{O}(MN\log N) = \mathcal{O}(BT\log T)$~\cite{NishantDiscreteRadar25}.

Let $G=ab=BT$ be the number of samples per frame and $\hat P$ the maximum
number of accepted paths, including the hypotheses of the accepted
ambiguous sets, which can be set by system design in relation to the number
of true targets $P$. Both proposed receivers compute one discrete Zak
transform per frame, with complexity $\mathcal{O}(G\log b)$ for frame~1 and
$\mathcal{O}(G\log a)$ for frame~2, i.e., $\mathcal{O}(G\log G)$ in total.

DS then performs the detection on the two grids, with
complexity $\mathcal{O}(G)$, and the matching, whose Hungarian algorithm
has complexity $\mathcal{O}(n^3)$ for a group of $n$ detections, i.e., at
most $\mathcal{O}(\hat P^3)$. The total complexity of DS is therefore
$\mathcal{C}_{\mathrm{DS}}=\mathcal{O}(G\log G+\hat P^3)$, where the
off-grid refinement, $\mathcal{O}(\hat P)$, is included. Since the groups
usually contain only a few detections, this is in practice
$\mathcal{O}(BT\log(BT))$, near-linear in the time--bandwidth product.

DS-XC additionally performs iterative cross-pilot cancellation and joint LS
gain estimation. Detection over up to $\hat P$ iterations costs
$\mathcal{O}(\hat PG\log G)$. Each LS update fits the gains of up to
$\hat P$ path templates, i.e., the unit-gain responses of the detected
paths, at a cost of $\mathcal{O}(G\hat P^2+\hat P^3)$, yielding a total LS
cost of $\mathcal{O}(G\hat P^3+\hat P^4)$ over the $\hat P$ iterations.
Thus,
$\mathcal{C}_{\mathrm{DS\text{-}XC}}=\mathcal{O}(\hat PG\log G+G\hat P^3+\hat P^4)$,
where both the construction of the $\hat P$ path templates,
$\mathcal{O}(\hat PG\log G)$, and the off-grid refinement,
$\mathcal{O}(\hat P)$, are included. This is
still linear in $G$ for a given number of accepted paths $\hat P$.
DS-XC is one possible receiver for the proposed design, and its complexity
can be reduced in several ways, e.g., by capping the number of accepted paths $\hat P$ by design, by
updating the LS solution recursively instead of refitting it at every
iteration, or by obtaining the comb templates directly from the analytical
DD response of the filter.
\begin{remark}[Integer shifts]
With integer delays and Doppler shifts and ideal pulse shaping, the comb
templates would be sparse and known in closed form. Thus, the cancellation
and the LS fit would involve only the peak and comb bins, and the
complexity would reduce to
$\mathcal{O}(G\log G+\hat PG+\rho\hat P^3+\hat P^4)$, which remains linear
in $G$, as in the fractional case.
\end{remark}

\begin{table}[!t]
  \centering
  \small
  \caption{Computational complexity, with $G = BT$ samples per frame and
  $\hat P$ accepted paths.}
  \label{tab:complexity}
  \begin{tabular}{@{}ll@{}}
    \toprule
    Method & Complexity \\
    \midrule
    Conventional radar & $\mathcal{O}(G^2)$ \\
    Zak-OTFS pulsone~\cite{NishantDiscreteRadar25} & $\mathcal{O}(G\log N)$ \\
    DS & $\mathcal{O}(G\log G+\hat P^3)$ \\
    DS-XC & $\mathcal{O}(\hat PG\log G+G\hat P^3+\hat P^4)$ \\
    DS-XC (integer shifts) & $\mathcal{O}(G\log G+\hat PG+\rho \hat P^3+\hat P^4)$\\
    \bottomrule
  \end{tabular}
\end{table}

\section{Baseline Systems}
\label{sec:compared}

We compare the proposed DS and DS-XC receivers with the following baselines. For a fair comparison, all systems use the system model of Section~\ref{sec:system model}, the Zak-OTFS transmitter and receiver of Fig.~\ref{fig:zak_blocks}, the same Gauss--sinc filters, and fractional delays and Doppler shifts. Their frame dimensions are listed in Table~\ref{tab:frames}.

\textbf{DD-Chirp~\cite{priya2024delay}:} A high-power DD pilot gives the aliased delays and the Doppler shifts. Paths with aliased delays are identified by the spreading of their pilot energy along the Doppler axis, caused by the pulses that fall outside the receive window, and their true delays are found by time-domain correlation with Doppler-matched dual chirps. Since the Doppler shifts are read directly from the DD grid, the Doppler spread must remain within the Doppler period, $2\nu_{\max} < \nu_p$.

\textbf{DD-Cosine~\cite{priya2026doppler}:} A high-power DD pilot gives the delays and the aliased Doppler shifts, and the spectrum of the echoes of a cosine pilot gives the true Doppler shifts, which are paired with the delays through their aliases and gains. Since the delays are read directly from the DD grid, the delay spread must remain within the delay period, $\tau_{\max} < \tau_p$.

\textbf{Single-frame Zak-OTFS~\cite{lin2026otfs}:} The standard DD sensing scheme, recently demonstrated in DISACT~\cite{lin2026otfs}, uses a single frame with a point pilot and operates up to the crystallization condition. It is processed with the same peak detection as the proposed receivers, and each detection is reported within the fundamental region of the frame. We consider a square frame with $M = N = \sqrt{G}$, a delay-covering frame, and a Doppler-covering frame.
\begin{table}[!t]
  \centering
\caption{Frame dimensions at the base point: nominal spread factor
  $\varsigma=4$, $G = 2^{12}\varsigma = 2^{14}$, on a simulated spread
  $0.9\varsigma=3.6$, i.e., delay and Doppler spreads of $243$ bins.}
  \label{tab:frames}
  \begin{tabular}{@{}lcc@{}}
    \toprule
    System & $M \times N$ & Energy split \\
    \midrule
    \textbf{Proposed (DS and DS-XC)} & $64\times256$ and $256\times64$ & $1/2$ per pilot \\
    
DD-Chirp~\cite{priya2024delay} & $64 \times 256$ & pilot + dual chirp \\
DD-Cosine~\cite{priya2026doppler} & $256 \times 64$ & pilot + cosine \\
    Single, square~\cite{lin2026otfs} & $128 \times 128$ & pilot \\
    Single, delay-cover~\cite{lin2026otfs} & $256 \times 64$ & pilot \\
    Single, Doppler-cover~\cite{lin2026otfs} & $64 \times 256$ & pilot \\
    \bottomrule
  \end{tabular}
\end{table}

\section{Simulation Results}
\label{sec:results}
\begin{figure*}[!t]
  \centering
  \subfloat[$P_d$ vs. $\varsigma$]{%
    \includegraphics[width=0.32\textwidth]{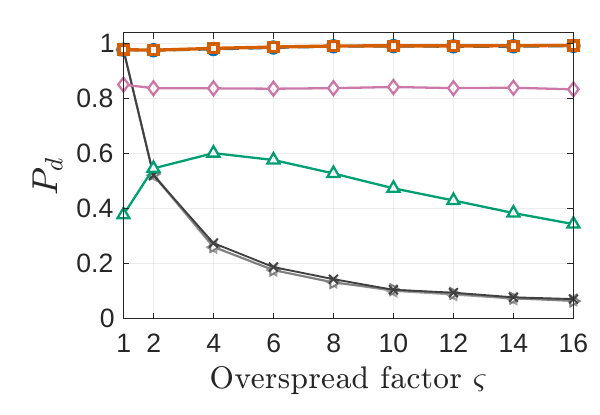}%
    \label{fig:pd_sigma}}\hfill
  \subfloat[$P_d$ vs. $P$]{%
    \includegraphics[width=0.32\textwidth]{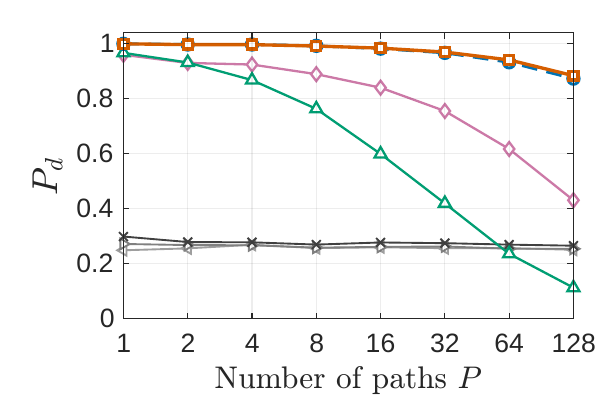}%
    \label{fig:pd_P}}\hfill
  \subfloat[$P_d$ vs. $\mathrm{SNR}_t$]{%
    \includegraphics[width=0.32\textwidth]{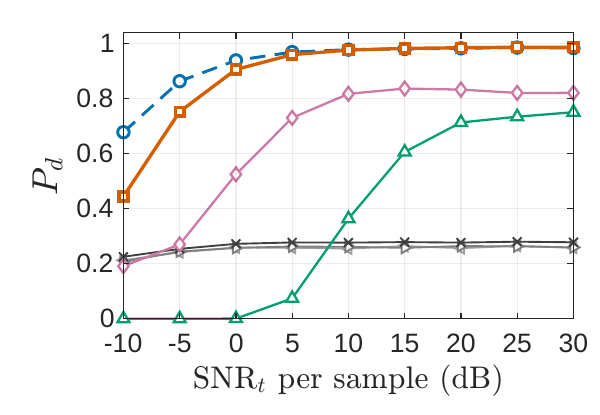}%
    \label{fig:pd_SNR}}\\
  \subfloat[$P_d$ vs. $\Delta\mathrm{SNR}$]{%
    \includegraphics[width=0.32\textwidth]{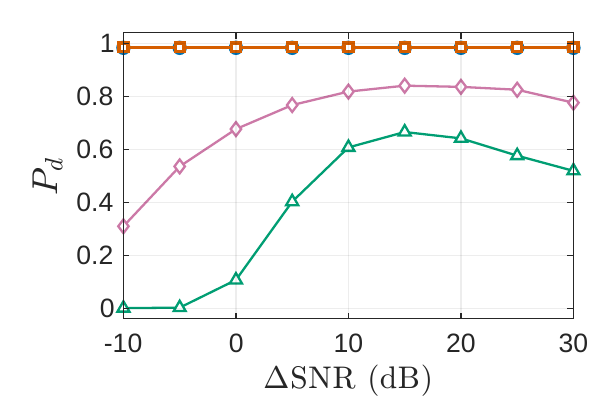}%
    \label{fig:pd_dSNR}}\hfill
  \subfloat[FA per realization vs. $\varsigma$]{%
    \includegraphics[width=0.32\textwidth]{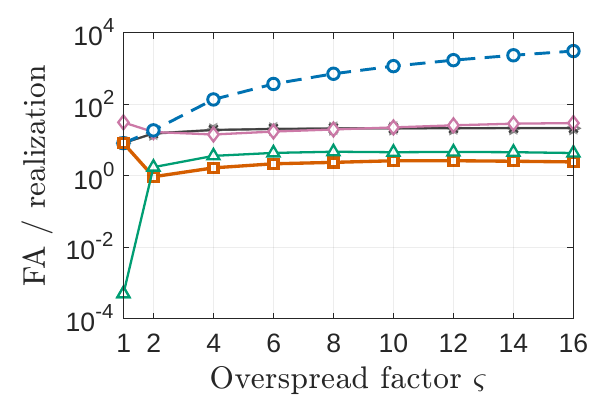}%
    \label{fig:fa_sigma}}\hfill
  \subfloat[FA per realization vs. $\Delta\mathrm{SNR}$]{%
    \includegraphics[width=0.32\textwidth]{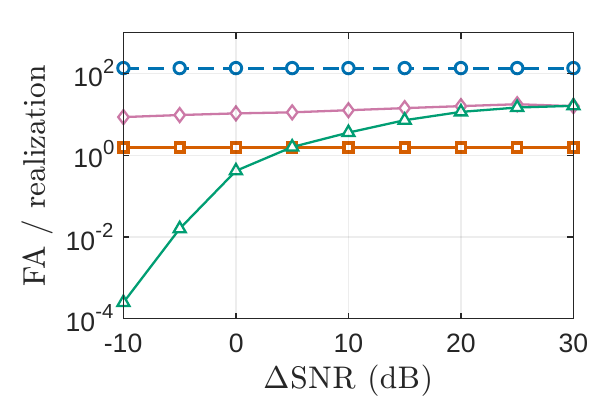}%
    \label{fig:fa_dSNR}}\\
  \subfloat[NMSE vs. $\mathrm{SNR}_t$]{%
    \includegraphics[width=0.32\textwidth]{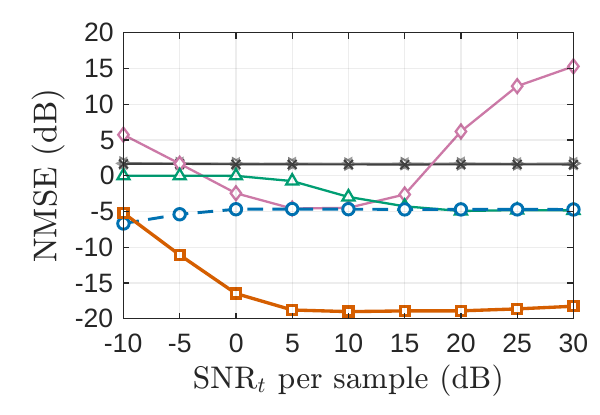}%
    \label{fig:nmse_SNR}}\hfill
  \subfloat[NMSE vs. $P$]{%
    \includegraphics[width=0.32\textwidth]{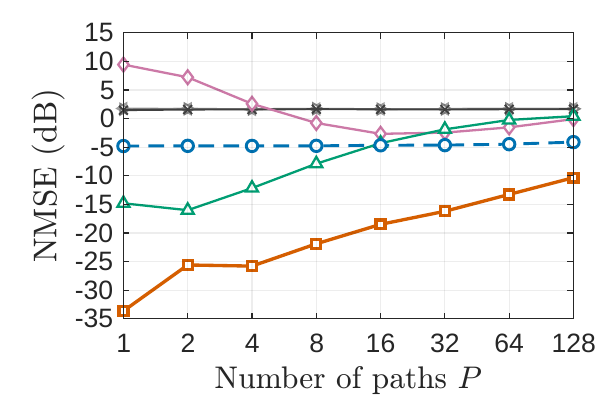}%
    \label{fig:nmse_P}}\hfill
  \subfloat[NMSE vs. $\Delta\mathrm{SNR}$]{%
    \includegraphics[width=0.32\textwidth]{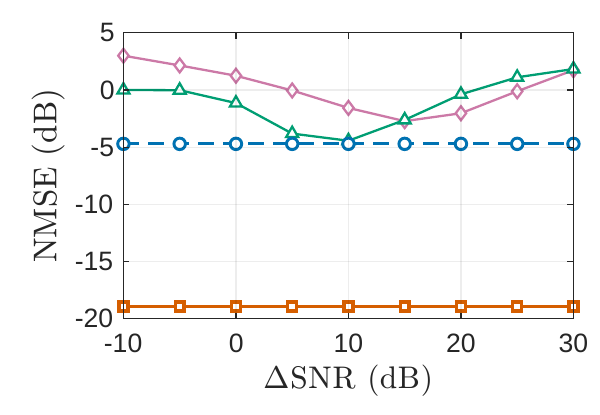}%
    \label{fig:nmse_dSNR}}

  \vspace{2mm}
  \resizebox{0.95\textwidth}{!}{
\usetikzlibrary{plotmarks}

\definecolor{cP24}{HTML}{009E73}
\definecolor{cP26}{HTML}{CC79A7}
\definecolor{cDS}{HTML}{0072B2}
\definecolor{cXC}{HTML}{D55E00}
\definecolor{cSD}{HTML}{808080}
\definecolor{cSN}{HTML}{A6A6A6}
\definecolor{cSQ}{HTML}{404040}
\begin{tikzpicture}[font=\footnotesize, line width=1.1pt]
  \newcommand{\lgd}[6]{%
    \draw[#3] (#1,#2) -- ++(0.9,0);
    \draw[#3] plot[mark=#4, mark size=2.6pt,
      mark options={solid, fill=white, #5}] coordinates {(#1+0.45,#2)};
    \node[anchor=west, inner sep=1pt] at (#1+1.0,#2) {#6};}
  \lgd{0}{0}     {cP24}{triangle}{}{DD-Chirp~\cite{priya2024delay}}
  \lgd{0}{-0.4}  {cSD}{triangle}{rotate=-90}{Single delay-cover~\cite{lin2026otfs}}
  \lgd{4.3}{0}   {cP26}{diamond}{}{DD-Cosine~\cite{priya2026doppler}}
  \lgd{4.3}{-0.4}{cSN}{triangle}{rotate=90}{Single Doppler-cover~\cite{lin2026otfs}}
  \lgd{8.9}{0}   {cDS, dash pattern=on 4pt off 3pt}{o}{}{Prop.~DS}
  \lgd{8.9}{-0.4}{cSQ}{x}{}{Single square~\cite{lin2026otfs}}
\lgd{12.7}{0}{cXC, line width=1.6pt}{square}{scale=0.8, fill=white}{Prop.~DS-XC}
  \draw[line width=0.5pt, black] (-0.15,0.25) rectangle (15.6,-0.65);
\end{tikzpicture}}
\caption{Sensing and channel estimation performance of the proposed
receivers and baselines. Unless swept, $\mathrm{SNR}_t=15$~dB, $G=2^{14}$,
$P=16$, $\varsigma=4$, and $N_{\mathrm{MC}}=1000$ Monte Carlo runs.
Detection probability $P_d$ versus (a)~$\varsigma$, with
$G=2^{12}\varsigma$, (b)~the path count $P$, (c)~$\mathrm{SNR}_t$, and
(d)~the power split $\Delta\mathrm{SNR}$ of DD-Chirp~\cite{priya2024delay}
and DD-Cosine~\cite{priya2026doppler}; false alarms per realization versus
(e)~$\varsigma$ and (f)~$\Delta\mathrm{SNR}$; channel NMSE versus
(g)~$\mathrm{SNR}_t$, (h)~$P$, and (i)~$\Delta\mathrm{SNR}$.}
  \label{fig:performance}
\end{figure*}
All systems transmit a single frame of $G = MN = BT$ samples without cyclic
prefix and observe the same channel and noise realizations. They therefore
share the same bandwidth $B$, frame duration $T = G/B$, and delay and Doppler
resolutions. The frame size grows with the spread factor as
$G = 2^{12}\varsigma$, and at each $G$ every system uses the frame dimensions
best suited to the channel spreads, as in Table~\ref{tab:frames}.
All systems transmit the same total energy $E_{\mathrm{tot}}$, measured after the transmit filter, and hence the same mean power $E_{\mathrm{tot}}/G$ per sample. We define the transmitted SNR per sample as $\mathrm{SNR}_t = E_{\mathrm{tot}}/(G\sigma_w^2)$, where $\sigma_w^2$ is the noise power per received sample defined in Section~\ref{sec:system model}. 

DD-Chirp and DD-Cosine are implemented as described in~\cite{priya2024delay,priya2026doppler}, with four changes that adapt them to our setting.
First, since the paths are fractional, each path spreads over neighboring DD bins. Only threshold crossings that are also local maxima are kept as candidates.
Second, the pilot-to-chirp power split $\Delta\mathrm{SNR} = \mathrm{SNR}_p - \mathrm{SNR}_c$ follows each paper's definitions: $\mathrm{SNR}_p = |x_p|^2/(N\sigma_w^2)$ for both methods, with $\mathrm{SNR}_c = 2A^2/\sigma_w^2$ for DD-Chirp and $\mathrm{SNR}_c = A^2/(N\sigma_w^2)$ for DD-Cosine, where $x_p$ is the pilot amplitude and $A$ the chirp or cosine amplitude. When the total energy is fixed, a larger $\Delta\mathrm{SNR}$ lowers $\mathrm{SNR}_c$. We therefore sweep $\Delta\mathrm{SNR}$ (Figs.~\ref{fig:pd_dSNR} and~\ref{fig:nmse_dSNR}) and use the best value of each method: $10$\,dB for DD-Chirp and $15$\,dB for DD-Cosine.
Third, the delays and Doppler shifts of the detected paths are refined to sub-bin accuracy, as for DS and DS-XC.
Finally,~\cite{priya2024delay} does not sweep $\mathrm{SNR}_t$, and its chirp-correlation threshold is set for a single chirp SNR ($\mathrm{SNR}_c = 23$\,dB). In our $\mathrm{SNR}_t$ sweep (Fig.~\ref{fig:nmse_SNR}), we therefore scale it in proportion to $\mathrm{SNR}_c$.

Let $S_\tau$ and $S_\nu$ denote the delay spread $\tau_{\max}$ and the Doppler spread $2\nu_{\max}$ in bins. For a nominal spread factor $\varsigma$, we set $S_\tau = S_\nu = \sqrt{0.9\,\varsigma G}$, so that the simulated channel has spread $0.9\,\varsigma$. Since the proposed frames cover a spread $\rho = \varsigma$, this margin keeps all paths, including the leakage of fractional paths into neighboring bins, within the covered region. The $P$ paths have fractional delays and Doppler shifts drawn uniformly within the spreads, and independent gains $h_i \sim \mathcal{CN}(0, 1/P)$, so that the expected channel power is one for any $P$.

The base point is $\varsigma = 4$, $P = 16$, $\mathrm{SNR}_t = 15$~dB, and
$G = 2^{14}$, and each sweep varies one of $\varsigma$, $\mathrm{SNR}_t$, and
$P$, with the other parameters fixed. In the sweep over $\varsigma$, the total
transmit energy $E_{\mathrm{tot}}$ is kept fixed, so that the peak SNR in the
DD grids is the same at every point, while $\mathrm{SNR}_t$ decreases as $G$
grows.
The detection threshold is set for $P_{\mathrm{FA}} = 10^{-2}$. A pairing is flagged as ambiguous when a swap increases the total cost by less than $\theta_g = 0.2$, and the stricter threshold $\theta_{xc} = 0.6$ is used inside the DS-XC cancellation loop. All systems use the same sub-bin refinement~\eqref{eq:refine}. All metrics are averaged over $N_{\mathrm{MC}} = 1000$ Monte Carlo
realizations per point.

\emph{Performance Metrics:}
A target is considered detected if its delay and Doppler errors are both below one bin. We report the detection rate $(P_d)$, the number of FA per realization, and the channel NMSE. For off-grid paths, the value at the peak bin underestimates the gain, so the output gains of all systems are corrected in the same way, using the pulse response at the estimated sub-bin position. Since the gains are estimated identically for all systems, the NMSE differences between them reflect detection and localization only: missed paths, FA, and wrong aliases all increase the NMSE. For the long-range drone scenario, we also report the range and speed errors, based on the generalized optimal sub-pattern assignment (GOSPA) metric~\cite{rahmathullah2017gospa} of order $2$. Detected targets contribute their actual errors, while every missed target and FA is penalized with the full range window and velocity span.

\subsubsection{Effect of the Spread Factor}
Fig.~\ref{fig:pd_sigma} shows that both proposed receivers keep a detection rate close to one at every spread factor, which confirms that the dual-frame design resolves overspread channels. The single-frame systems, as in DISACT~\cite{lin2026otfs}, perform well only at $\varsigma = 1$, where the proposed frames reduce to a single square frame. As $\varsigma$ grows, an increasing fraction of the paths falls outside their fundamental region and is reported at a wrong alias, so their detection rate drops below $0.1$ at $\varsigma = 16$. DD-Cosine and DD-Chirp each resolve the aliasing in only one dimension, and their detection rates remain well below those of the proposed receivers over the whole range.
The two proposed receivers differ in their FA, shown in Fig.~\ref{fig:fa_sigma}. Since $a = \sqrt{G/\rho}$ is fixed and $\rho = \varsigma$ in this sweep, each path in DS produces up to $(\rho - 1)^2$ pairings of comb peaks, so the FA of DS grow rapidly with $\varsigma$. For $\varsigma \le 2$, however, the FA of DS remain comparable to those of the baselines. DS is therefore a valid low-complexity option for mildly overspread channels. By canceling the combs, DS-XC reduces the FA below those of all other systems and achieves the lowest NMSE at every spread factor, whereas the NMSE of DS degrades as the comb ghosts grow with $\varsigma$.

\subsubsection{Effect of the SNR}
Fig.~\ref{fig:pd_SNR} shows that the proposed receivers reach a detection rate close to one already at $\mathrm{SNR}_t \approx 0$~dB. DS achieves a higher detection rate than DS-XC at low SNR, since DS-XC accepts a target only after its pairing becomes reliable. Both DD-Chirp and DD-Cosine perform poorly at low SNR, since they rely on a high-power pilot, and DD-Chirp is the most affected. The NMSE in Fig.~\ref{fig:nmse_SNR} confirms the advantage of the proposed design: the NMSE of DS is limited by its FA, whereas the NMSE of DS-XC keeps decreasing with the SNR and remains the lowest among all systems.

\subsubsection{Effect of the Number of Paths}
Fig.~\ref{fig:pd_P} shows that the detection rate of the proposed receivers decreases only slowly with the number of paths, and remains close to $0.9$ even for $P = 128$. As $P$ grows, more targets share a projection bin, as predicted by~\eqref{eq:prob_basic}, but their different gains still allow most of them to be paired correctly. The detection rate of DD-Cosine and DD-Chirp drops much faster, since their pairing and refinement steps are more sensitive to overlapping paths. The low FA of DD-Chirp in Fig.~\ref{fig:fa_sigma} must be read together with its low detection rate: its first-stage test retains only a few strong paths, which are almost always correct.

\begin{table}[!t]
  \centering
  \small
  \setlength{\tabcolsep}{4pt}
\caption{Long-range drone scenario with the base-point frames, $P = 16$, $\mathrm{SNR}_t = 15$~dB, and $N_{\mathrm{MC}} = 1000$: parameters (top) and results (bottom), with $e_r$ and $e_v$ the GOSPA range and speed errors.}
  \label{tab:scenario}
  \begin{tabular}{ll}
  Parameters & Values\\
    \toprule
    Range window / max.\ speed & $100$~km / $\pm 69$~m/s \\
    Carrier $f_c$ / bandwidth $B$ & $5.9$~GHz / $364$~kHz \\
    Frame size $G$ / duration $G/B$ & $2^{14}$ / $45$~ms \\
    Spreads $\tau_{\max}$ / $2\nu_{\max}$ & $667~\mu$s / $5.40$~kHz \\
    Spread factor $\varsigma = 2\tau_{\max}\nu_{\max}$ & $3.6$ \\
    frame~1, $64\times256$: $\tau_{p1}$ / $\nu_{p1}$ & $176~\mu$s / $5.69$~kHz \\
    frame~2, $256\times64$: $\tau_{p2}$ / $\nu_{p2}$ & $703~\mu$s / $1.42$~kHz \\
    Covered spread $\tau_{p2}\nu_{p1}$ & $\rho = 4$ \\
    Range bin / velocity bin & $412$~m / $0.56$~m/s \\
    \bottomrule
  \end{tabular}

  \vspace{2mm}

\begin{tabular}{lccccc}
    \toprule
    System & $P_d$ & FA & $e_r$ (km) & $e_v$ (m/s) & NMSE (dB) \\
    \midrule
    \textbf{Prop.\ DS-XC} & $\mathbf{0.98}$ & $\mathbf{1.7}$ & $\mathbf{25.0}$ & $\mathbf{34.3}$ & $\mathbf{-18.6}$ \\
    Prop.\ DS      & $0.98$ & $136.6$ & $207$  & $284$  & $-4.7$ \\
    DD-Chirp~\cite{priya2024delay}    & $0.50$ & $2.2$  & $56.5$ & $77.4$ & $-4.4$ \\
    DD-Cosine~\cite{priya2026doppler} & $0.81$ & $15.5$ & $76.2$ & $105$  & $0.9$ \\
    Single, square & $0.28$ & $19.1$  & $98.0$ & $134$  & $1.6$ \\
    \bottomrule
  \end{tabular}
\end{table}
\subsubsection*{Long-Range Drone Scenario}
Table~\ref{tab:scenario} considers drones within a $100$~km range window and radial speeds up to $\pm 69$~m/s, at a carrier frequency of $5.9$~GHz, which corresponds to a spread factor of $3.6$. Neither a single frame nor the baselines can cover both spreads, whereas the proposed frames cover a spread $\rho = 4$. DS-XC detects $98\%$ of the targets, compared with $81\%$ for DD-Cosine and $28\%$ for the square single frame, and achieves the lowest range and speed errors and an NMSE of $-18.6$~dB. DS reaches the same detection rate, but its FA due to the cross-pilot interference dominates its range and speed errors.

\subsubsection*{Practical Considerations}
The PAPR of the point pilots considered in this paper equals $M$, e.g.,
$21.1$~dB for a $128\times128$ frame. Radar systems, however, require a low
PAPR, and therefore usually spread the pilot with a constant-amplitude
zero-autocorrelation (CAZAC) sequence, whose PAPR is close to $0$~dB.
 A natural choice is a Zadoff--Chu (ZC) sequence of length $G$~\cite{chu1972polyphase,andrews2022zc}, $x_u[n] = e^{j\pi u n (n + c)/G}$, with $c = G \bmod 2$ and $\gcd(u, G) = 1$. The proposed design also applies to frames with spread pilots. With ZC spreading, the aliasing of each frame is sheared in the DD lattice, so the shape of the
projection grid and the pairing change accordingly. Moreover, using
orthogonal or nearly orthogonal bases, e.g., ZC sequences with coprime
parameters, could reduce the comb interference of Fig.~\ref{fig:comb},
possibly making the cancellation of DS-XC unnecessary. The verification of
these properties with fractional paths and practical filters is left for
future work.

The proposed design can in principle also carry data: one frame, e.g.,
frame~1, can carry data superimposed on its point pilot, while frame~2
carries only its pilot. Data transmission over overspread channels,
however, requires dedicated transmission strategies and receiver
algorithms, which are beyond the scope of this paper.

\section{Conclusion}
\label{sec:conc}

We proposed a frame design that enables DD sensing of fractional overspread channels, which arise in long-range mobility scenarios beyond the crystallization limit of a single DD frame. The design combines frames with different delay and Doppler periods, which alias each path differently, so that their joint observation identifies the path over a region larger than any single frame. We showed that two frames with swapped dimensions recover each target by pairing their detections through a common projection and their gains, and generalized the design to frames with non-divisible dimensions, which enlarge the unaliased region and reduce the pilot PAPR. We implemented the design with two superimposed Zak-OTFS frames and proposed two receivers. DS offers a low-complexity solution for mildly overspread channels, while DS-XC cancels the cross-pilot interference and outperforms all baselines in detection rate, false alarms, and channel NMSE, with a complexity that grows near-linearly with the time--bandwidth product. In a long-range drone scenario, DS-XC detects $98\%$ of the targets, compared with $81\%$ for the best baseline.

Future work includes pilot spreading to reduce the PAPR and the cross-pilot
interference, data transmission, receivers with lower complexity and higher
accuracy, the application to other DD modulations, and experimental
validation in aerial, satellite, and underwater scenarios.

\balance

\bibliographystyle{IEEEtran}
\bibliography{IEEEabrv,biblio}

\end{document}